\documentclass[fleqn,usenatbib]{mnras}
\usepackage[T1]{fontenc}
\usepackage{ae,aecompl}
\usepackage{graphicx}	
\usepackage{amsmath}	
\usepackage{amssymb}
\usepackage{pdflscape}
\usepackage{siunitx}
\usepackage{dcolumn}
\usepackage{multicol} 
\usepackage{hyperref}

\title[UV luminosity function of Galaxies]{{Galaxy luminosity functions at redshifts $0.6-1.2$} in the Chandra Deep Field South}

\author[M. Sharma et al.]
      {{M. Sharma$^{1}$\thanks{E-mail: mnushv@gmail.com (MS)},
          M. J. Page$^{1}$, A. A. Breeveld$^{1}$}\\
         $^1$Mullard Space Science Laboratory,
         University College London,
         Holmbury St Mary, Dorking, Surrey, RH5 6NT, UK}
         
\date{Accepted 2022 January 14; Received 2021 December 22; in original form 2021 October 28}
       
\pubyear{2022}

\begin{document}
\label{firstpage}
\pagerange{\pageref{firstpage}--\pageref{lastpage}}
\maketitle

\begin{abstract}
We present the rest-frame  Ultra$-$Violet (UV) galaxy luminosity 
function (LF) and luminosity density (LD)
measurements in the far-UV (1500 {\AA}) wavelength, in the redshift
range $z = 0.6 - 1.2$. 
The UV LF is derived using \textit{XMM-Newton} Optical Monitor
(XMM-OM), ultraviolet ($1600-4000$ {\AA}) observations of the
Chandra Deep Field South, over an area of 396
arcmin\textsuperscript{2}. Using the deep UV imaging of the 
CDFS, we identified $> 2500$ galaxies in our sample with 
UVW1$\mathrm{_{AB}} \leq 24.5$ mag.
This sample along with various other catalogues containing redshift
information, is used to calculate the binned representation of the
galaxy UV LF in the two redshift bins $0.6 \leq z < 0.8$ and 
$0.8 \leq z < 1.2$, having a wide range of 1500 {\AA} rest-frame
UV magnitudes ($\Delta M_{1500} \simeq 3$), reaching $\simeq 1-1.5$
magnitudes fainter than previous studies at similar redshifts. The binned
LF is described well by the Schechter function form. Using 
maximum-likelihood the Schechter function is fitted to the unbinned 
data to obtain the best-fit values of the the UV galaxy LF parameters.
We find that characteristic magnitude $M^*$ brightens by $0.8$ mag from 
$ z = 0.7$ to $z = 1$, implying an increase in the star 
formation activity between these redshifts, as reported by past 
studies. 
Our estimate of the faint-end slope $-1.10^{+0.19}_{-0.18}$ is 
on the shallower side compared with previous studies at $ z = 0.7$, 
whereas a value of $-1.56^{+0.19}_{-0.18}$ estimated for 
$ z = 1.0$, agrees with previous results given the uncertainties.

\end{abstract}


\section{Introduction}
\label{sec:1}
Luminosity is one of the characteristic global properties of a 
galaxy, mainly controlled by other important quantities (e.g. 
the total mass, gas mass, stellar population etc. in the galaxy). 
Using multi-wavelength imaging surveys along with spectroscopic or 
photometric redshifts, luminosity can be estimated at various 
rest-frame wavelengths. These estimates can be used to produce a 
one-point statistic called the luminosity function (LF) $-$ 
the number density of galaxies as a function of luminosity.
The LF can be used to obtain a relative distribution of power 
(produced at
a given rest-frame wavelength) among galaxies of different luminosities 
and masses at a particular average epoch in the history of the Universe.

The observed functional form of galaxy LFs is different from what is 
predicted from the ${\Lambda}$-CDM model (assuming a mass to light ratio),
both in normalisation and the shape at the bright and faint ends
\citep[e.g.][]{2001MNRAS.326..255C,2003MNRAS.339.1057Y,2010ApJ...717..379B}.
This observation has led to a suggestion that there must be additional 
physical processes involved, causing moderation of the star formation at 
high and low luminosity regimes through baryonic feedback mechanisms 
\citep{2010ApJ...710..903M}. 
In particular, stellar feedback - outflows from supernovae (SNe) only \citep[e.g.][]{1986ApJ...303...39D,1991ApJ...379...52W}
or winds from hot young stars in addition to SNe \citep[e.g.][]{2003MNRAS.339..312S,2006MNRAS.373.1265O,
    2012MNRAS.421.3522H,2014MNRAS.445..581H,2017MNRAS.466.1903G} 
are thought to dictate the shape at low-luminosities, whereas feedback from 
active galactic nuclei - through relativistic outflows countering the 
cooling flows in cluster atmospheres \citep[e.g.][]{1998A&A...331L...1S,
    2003ApJ...596L..27K,2005ApJ...618..569M,2006MNRAS.365...11C,
    2006MNRAS.370..645B}
are thought to be responsible for the LF shape at high luminosities 
\citep[see][for reviews]{2009Natur.460..213C,2012ARA&A..50..455F,
    2018NatAs...2..198H}.
It is clear that the functional form of the galaxy LF encodes information 
about these physical processes (i.e. feedback, gas accretion, mergers, etc.) 
dictating galaxy formation and evolution 
\citep[e.g.][and references therein]{2000MNRAS.319..168C,2003ApJ...599...38B,
    2015ARA&A..53...51S}.
So, constraining this functional form at different redshifts is very
important to gain insights into the properties of the galaxy populations at those redshifts. The evolution of the LF parameters with redshift can be used to 
understand the evolutionary history of a class of galaxies under 
consideration.

A single-aged star-forming population produces half 
of its bolometric radiation during the first 1 per cent of its total 
lifetime \citep[see e.g.][]{1999ApJS..123....3L}. Its bolometric 
luminosity is dominated by UV emission, 
coming from young and massive stars with very short lifetimes 
\citep[see][]{2014ARA&A..52..415M}.
This instantaneous star-formation feedback regulates the faint-end
slope of the UV LF, as it occurs, unlike other longer wavebands, where the
LFs are shaped by accumulative feedback taking place over very long 
time scales.
Therefore, rest-frame UV is considered an important tracer for 
instantaneous star-formation rate density (SFRD), although affected by dust \citep[see][]{2012ARA&A..50..531K}. This is also one of the reasons 
why the LF in the rest-frame UV has emerged as a key 
diagnostic for studying the evolution of galaxies and contribution of these
galaxies towards cosmic star formation rate density \citep[see][]{2014ARA&A..52..415M}.

In the recent history of the field, many studies used data from 
spaceborne and/or ground based observatories to produce estimates of 
the galaxy LF in different redshift ranges (up to $z = 10$). 
Some of these works use drop-out techniques along with various colour selection
criterion to identify the sources and estimate photometric 
redshifts
\citep[e.g.][]{2006ApJ...653..988Y,2007ApJ...654..172D,2009ApJ...706.1136O,2010ApJ...725L.150O,2015ApJ...803...34B,2015MNRAS.452.1817B,2017ApJ...838...29M,2017ApJ...851...43S}, while
others derive their photometry in the rest-frame band of interest 
using model fits to the spectral energy distribution (SED) by combining
photometry in different wave bands
\citep[e.g.][]{2004A&A...421...41G,2007ApJ...654..172D,
    2015MNRAS.452.1817B,2016MNRAS.456.3194P,refId0,2020MNRAS.494.1894M}.
Some studies use their photometry with redshifts from other surveys 
\citep[e.g.][]{2005ApJ...619L..43A,2005ApJ...619L..15W,
    2010ApJ...720.1708H,2015ApJ...808..178H}.
Recently some works have tried to use gravitational lensing to get to 
extremely faint absolute magnitudes
\citep[e.g.][]{2016ApJ...832...56A,2019MNRAS.486.3805B}.

In the redshift range ($z \leq 1.5$), the UV LF can be estimated from
data obtained from the 
\textit{GALEX} \cite[\textit{Galaxy Evolution Explorer};][]{2005ApJ...619L...1M}
satellite.
Using rest-frame 1500 {\AA} broad-band photometry from the \textit{GALEX} 
FUV filter ($1750-2800$ {\AA}), the UV LF
parameters were obtained at low redshifts by 
\citet{2005ApJ...619L..15W} and \citet{2005ApJ...619L..31B}.
The NUV filter on \textit{GALEX} was used for studying the UV LF of galaxies at 
$0.2 \leq z \leq 1.2$ in \citet{2005ApJ...619L..43A}. 
A relatively flat faint-end slope (see eq. \ref{eqn:phi}), 
$\alpha \simeq -1.2$, was reported
by these works at low redshifts $(z < 0.4)$. However, they
calculated a steep and almost constant value, $\alpha \simeq -1.6$,
in the redshift range $0.6 \leq z \leq 1.2$ 
\citep[e.g.][]{2005ApJ...619L..43A, 2005ApJ...619L..15W}.
Using the Early Release Science data of the 
GOODS - South field \citep[Great Observatories Origins Deep Survey;][]{2004ApJ...600L..93G} from the WFC3/UVIS instrument on board 
the \textit{Hubble Space Telescope (HST)}, 
\citet{2010ApJ...725L.150O} obtained the faint-end slope,
$\alpha = -1.52 \pm 0.25$ in the redshift interval $0.5 < z < 1$.
\citet{2015ApJ...808..178H} have calculated the UV LF in the 
redshift range $0.2 < z < 1.2$, using data from UV/Optical Telescope 
\citep[UVOT;][]{2005SSRv..120...95R} on-board the \textit{Neil Gehrels Swift Observatory} \citep[][]{2004ApJ...611.1005G}, in the 
Chandra Deep Field South \citep[CDFS;][]{2001ApJ...551..624G}. They have used U-band selection with UV 
photometry at shorter wavelengths. Their values of the 
UV LF parameters are within $1 \sigma$ of \citet{2005ApJ...619L..43A}.
\citet{2021MNRAS.506..473P} used the UVW1 filter of the Optical/UV
telescope, the \textit{XMM-Newton} Optical Monitor (XMM-OM; \citet{2001A&A...365L..36M}) on-board the \textit{XMM-Newton}
observatory, and derived the 1500 {\AA} UV LF from observations of the 13 Hr field.

Some authors have derived UV LFs at $z < 1.2$ by selecting galaxies at longer
wavelengths and extrapolating their SEDs to rest-frame FUV.
In \citet{2004A&A...421...41G}, the $I$-band selected data set from
the FORS Deep Field are employed to calculate a range of values
$-1.14 < \alpha < -0.96$ for redshifts $0.45 < z < 1.1$. 
\citet{refId0} have used ${I}_{AB}$-band selected data and measured 
$\alpha$ to be in the range $-1.17 \pm 0.05$ to $-0.91 \pm 0.16$ for the redshift
range $0.3 < z < 1.1$ using the VVDS Survey.
Another recent work covering our redshift range is \citet{2020MNRAS.494.1894M}.
They used a large data-set from \textit{CFHT} Large Area U-band Deep Survey 
\citep[CLAUDS;][]{2019MNRAS.489.5202S}
and HyperSuprime-Cam Subaru Strategic Program \citep[HSC-SSP;][]{2018PASJ...70S...4A} for their LF estimates.
They also re-analysed the \textit{GALEX} data to measure the UV LF at redshift
$z < 0.9$.
The above mentioned three works calculate galaxy LFs to higher redshifts and 
we only quote their results 
falling within the redshift range of our study (i.e. $z < 1.2$).

In this paper we use observations of the Chandra Deep Field South
(CDFS) from the UVW1 filter on XMM-OM. 
The UVW1 filter on the XMM-OM has a central wavelength 2900 {\AA}
and samples the 1500 {\AA} flux better than \textit{GALEX} at 
redshift 
$z = 1$ where the \textit{GALEX} NUV pass-band is too blue, and the rest 
frame 
1500 {\AA} UV radiation is placed at the tail of the filter. 
In addition to that, the full width at half-maximum (FWHM) of the point 
spread 
function (PSF) for \textit{GALEX} NUV filter is $\simeq 5$ arcsec \citep{2007ApJS..173..682M}.
This big PSF puts its data at higher risk to source confusion 
as compared to UVW1 data obtained at a sharper PSF of 
$\simeq 2$\footnote{\url{https://xmm-tools.cosmos.esa.int/external/xmm_user_support/documentation/uhb/XMM_UHB.html}} arcsec.
\textit{HST} solves this problem as it has excellent
resolution, but covers a very small sky area, so its data might
be prone to cosmic variance.
So, in this redshift range, XMM-OM can be used to put better constraints on 
the UV LF parameters over larger fields than covered by the \textit{HST}
UV instrumentation and at a better spatial resolution compared with
\textit{GALEX}.
We mainly compare our results to \citet{2005ApJ...619L..43A,2015ApJ...808..178H} and \citet{2021MNRAS.506..473P} as they 
use the similar redshift binning scheme and direct rest-frame FUV observations for their LF estimates.

The rest of this paper is structured as follows. The observations, 
data and the reduction process are explained in section \ref{sec:2}. 
In section \ref{sec:3} we discuss the completeness simulations, possible 
biases and source-confusion.
The cross-correlations to the ancillary data, which include the 
spectroscopic and
photometric redshifts are used to identify the sources in 
section \ref{sec:4}.
Various corrections are made to the data in section \ref{sec:5}, before final analysis.
In the following section \ref{sec:6}, we discuss the 
methods to construct the UV LF and derive the LF parameters. Then, 
in section \ref{sec:7}, we mention the possible impacts of cosmic variance
on our calculations. Section \ref{sec:8} contains the calculations for the luminosity density.
The results are presented in section \ref{sec:9},
and their implications are discussed in section \ref{sec:10}. 
Finally, we conclude this paper in section \ref{sec:11}.
Throughout the paper we adopt a flat cosmology with
$\Omega_{\Lambda}=0.7$, $\Omega_{M}=0.3$ and Hubble's constant 
$H_0=70$\,km\,s$^{-1}$\,Mpc$^{-1}$. The distances (and volumes) 
are calculated in comoving co-ordinates in Mpc (and Mpc$^{3}$).
We use the AB system of magnitudes \citep{1983ApJ...266..713O}.

\section{Observations \& Data}
\label{sec:2}
\subsection{XMM-OM observations of the CDFS}
\label{sec:2.1}

\begin{table*}
  \centering
  \caption{Here we tabulate the UVW1 observations of the CDFS from XMM-OM. 
  First two columns are observation ID (OBS ID), Epoch (Observation dates). Third column lists the configurations in which the observation were taken (i.e. whether `full-frame' (\texttt{ff}), `rudi5' (\texttt{r5}) or both (\texttt{r5 + ff}) as explained in section \ref{sec:2.1}). 
  The following columns represent the pointing sky 
  co-ordinates RA and DEC (in degrees) and the Exposure time (Exp. time) in kilo-seconds.}
  \label{tab:obs}
  \begin{tabular}{llcccr}
        \hline\hline
        \noalign{\vskip 0.5mm}
        OBS ID  & 
        Epoch  &  
        Configuration &
        RA  &  
        DEC  &  
        Exp. time \\
        &   
        &
        &  
        (deg)
        &  
        (deg)  
        &  
        (ks) \\
        \hline
        \noalign{\vskip 0.5mm}
        0108060501 & 2001-07-28 & \texttt{r5} & 03h 32m 29.29s & -27d 48' 40.0" & 1.48 \\
        0108060601 & 2002-01-13 & \texttt{ff} & 03h 32m 27.99s & -27d 48' 50.0" & 5.0 \\
        0108060701 & 2002-01-15 & \texttt{r5 + ff}  & 03h 32m 26.70s & -27d 48' 40.0" & 10.0 \\
        0108061801 & 2002-01-16 & \texttt{ff}  & 03h 32m 27.99s & -27d 48' 30.0" & 5.0 \\
        0108061901 & 2002-01-17 & \texttt{ff}  & 03h 32m 27.99s & -27d 48' 10.0" & 5.0 \\
        0108062101 & 2002-01-20 & \texttt{ff}  & 03h 32m 29.29s & -27d 48' 20.0" & 5.0 \\
        0108062301 & 2002-01-23 & \texttt{ff}  & 03h 32m 27.99s & -27d 48' 10.0" & 5.0 \\
        \\
        0555780101 & 2008-07-05/06 & \texttt{r5} & 03h 32m 42.29s & -27d 45' 05.0" &
        8.8 \\
        0555780201 & 2008-07-07/08 & \texttt{r5} & 03h 32m 42.29s & -27d 45' 35.0" &
        8.8 \\
        0555780301 & 2008-07-09/10 & \texttt{r5} & 03h 32m 39.99s & -27d 45' 35.0" &
        8.5 \\
        0555780401 & 2008-07-11/12 & \texttt{r5} & 03h 32m 39.99s & -27d 45' 05.0" &
        8.8 \\
        \\
        0555780501 & 2009-01-07 & ff & 03h 32m 25.00s & -27d 49' 25.0" & 4.0 \\
        0555780601 & 2009-01-11 & ff & 03h 32m 25.00s & -27d 49' 55.0" & 4.0 \\
        0555780701 & 2009-01-13 & ff & 03h 32m 25.00s & -27d 50' 25.0" & 4.0 \\
        0555780801 & 2009-01-17 & ff & 03h 32m 22.70s & -27d 49' 25.0" & 4.0 \\
        0555780901 & 2009-01-19 & ff & 03h 32m 22.70s & -27d 49' 55.0" & 4.0 \\
        0555781001 & 2009-01-23 & ff & 03h 32m 22.70s & -27d 50' 25.0" & 3.5 \\
        0555782301 & 2009-01-25 & ff & 03h 32m 22.70s & -27d 50' 25.0" & 3.5 \\
        \\        
        0604960301 & 2009-07-05/06 & \texttt{r5} & 03h 32m 42.29s & -27d 46' 09.0" & 10.0 \\ 
        0604960201 & 2009-07-14/18 & \texttt{r5} & 03h 32m 39.99s & -27d 45' 35.0" & 7.12\\
        0604960101 & 2009-07-27/28 & \texttt{r5} & 03h 32m 42.29s & -27d 45' 35.0" & 10.0 \\
        0604960401 & 2009-07-29/30 & \texttt{r5} & 03h 32m 39.99s & -27d 46' 07.0" & 7.0 \\
        0604961101 & 2010-01-05 & \texttt{r5} & 03h 32m 25.00s & -27d 48' 52.3" & 10.0 \\
        0604961201 & 2010-01-09 & \texttt{r5} & 03h 32m 22.30s & -27d 46' 09.0" & 5.0 \\
        0604960701 & 2010-01-13 & \texttt{r5} & 03h 32m 22.70s & -27d 49' 25.0" & 10.0 \\
        0604961301 & 2010-01-20 & \texttt{r5} & 03h 32m 25.00s & -27d 49' 25.0" & 2.6 \\
        0604960601 & 2010-01-27 & \texttt{r5} & 03h 32m 25.00s & -27d 49' 55.0" & 10.0 \\
        0604960801 & 2010-02-06 & \texttt{r5} & 03h 32m 22.30s & -27d 50' 31.0" & 8.1 \\
        0604961001 & 2010-02-14 & \texttt{r5} & 03h 32m 22.70s & -27d 49' 55.0" & 9.0 \\
        0604961801 & 2010-02-18 & \texttt{r5} & 03h 32m 22.59s & -27d 49' 34.7" & 9.0 \\
        \hline
  \end{tabular}\\
\end{table*}

The Chandra Deep Field South (CDFS) : one of the deepest 
observation sites of the Chandra X-ray Observatory \citep{2008ApJS..179...19L}, 
has been a target of many space and ground-based observation campaigns in different wavebands. It is centred on 
RA 3h 32m 28.0s DEC $-27^{\circ} 48' 30"$ (J2000.0) \citep{2002ApJ...566..667R}.
This field is popular for having very low HI column density
\citep{1992ApJS...79...77S}, hence allows very deep observations.

The main aim of the \textit{XMM-Newton} survey in the Chandra Deep Field South 
(XMM-CDFS) was to detect and study the X-ray spectral properties of obscured
X-ray sources. The field was observed for more than 3 Ms to acquire good 
quality X-ray spectra for heavily obscured AGN, at all possible redshifts 
\citep{2011A&A...526L...9C}.

In addition to the X-ray instruments, the \textit{XMM-Newton} observatory carries the XMM-OM, an optical/UV telescope with a
primary task to complement the X-ray observations. It is a 
very capable stand-alone instrument with a wavelength coverage in the
range $180-590$ nm over six broadband filters, a fine PSF $\simeq 2"$ (FWHM) 
for each filter, over the entire field of view (FOV) spanning 
$17\times17$ arcmin\textsuperscript{2} 
\citep[for more details see][]{2001A&A...365L..36M}. We refer to Table 1 and 
Fig. 1 in \citet{2012MNRAS.426..903P} for other properties 
of the XMM-OM filter pass-bands.
More details about the different image configurations can be found in the
\textit{XMM-Newton} User's Handbook
\footnote{\url{https://xmm-tools.cosmos.esa.int/external/xmm_user_support/documentation/uhb/XMM_UHB.html}}.
XMM-OM can be used to take observations in the optical/UV 
regimes and build source catalogues in most fields observed by \textit{XMM-Newton}.
We use the data from the UVW1 filter, which is the most relevant to
derive the 1500 {\AA} LF at the redshifts we are interested in.
Our catalogue of the XMM-CDFS Deep Survey consists of 30 \textit{XMM-Newton}
observations in six different
observational epochs between July 2001 and February 2010. There are 11
observations corresponding to 2 epochs, acquired with the XMM-OM in `full frame
low-resolution' configuration to obtain single exposures of the whole
FOV in $2\times2$ binned resolution giving pixel size of 0.95 arcsec
\citep{2001A&A...365L..36M}.
These full-frame images have exposure durations of 5 ks each.
There are 17 other observations spanning 4 epochs that were acquired in
`default' (also known as `rudi-5') configuration, where a setup of 5 
consecutive exposures is used to cover 92 per cent of the FOV with the same binning and 
pixel scale as full-frame mode. In the rudi-5 exposures, UVW1 observations
have an exposure duration ranging from $0.8 - 5$ ks. One observation contains exposures
in both configurations \citep[see Table \ref{tab:obs} and also][for details of observations]{2015A&A...574A..49A}.

\subsection{XMM-OM data reduction and processing}
\label{sec:2.2}

\begin{figure}
  \centering
  \includegraphics[width=\columnwidth]{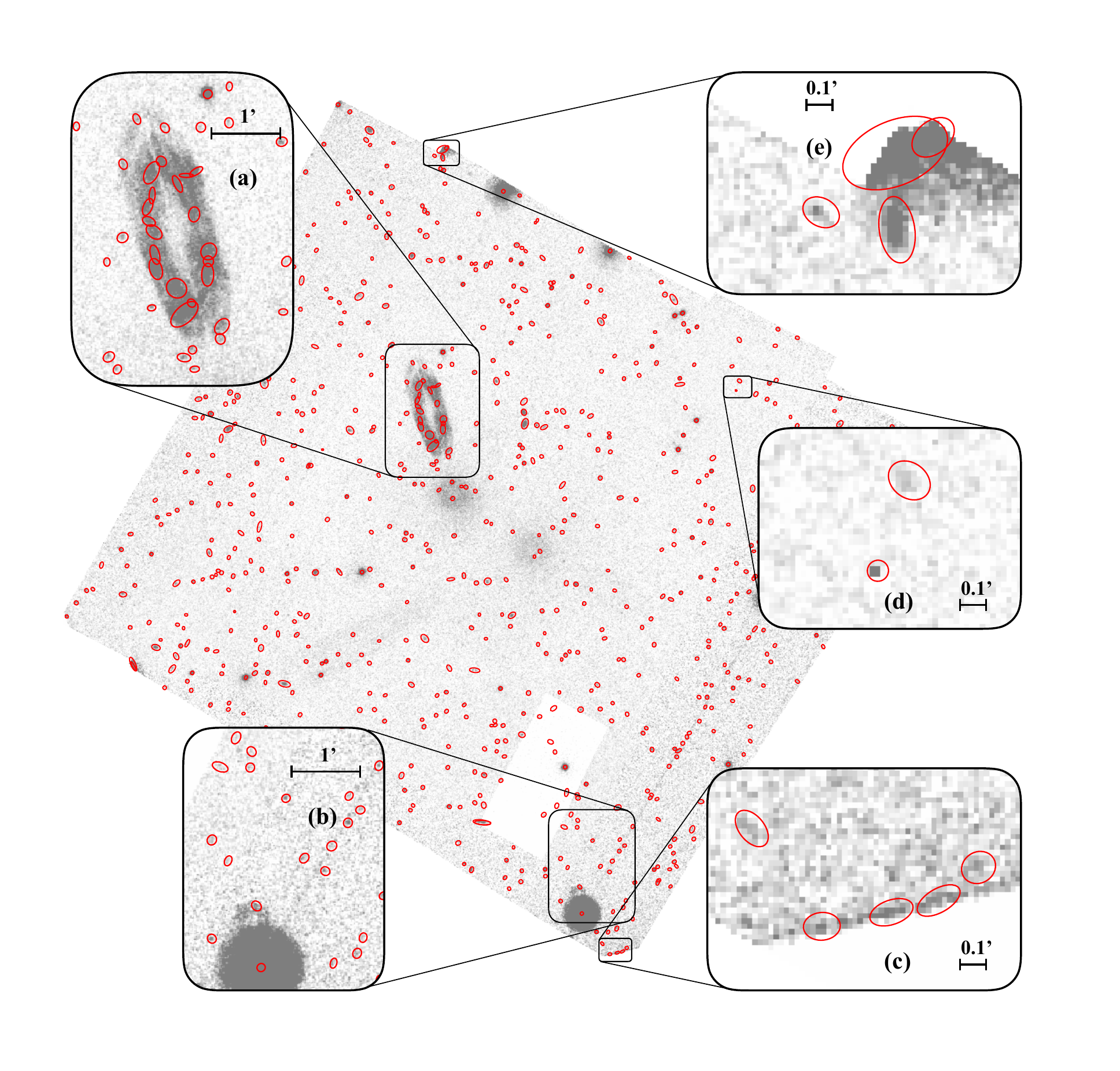}
  \caption{
  Example image created by mosaicing two different 
  exposures of UVW1 images of the CDFS. These exposures are processed
  using the standard SAS tasks without any additional processing.
  All the detections by \textsc{omdetect} are shown by red ellipses.
  At the center of the image the scattered background light structures 
  (explained in section \ref{sec:2.2}) are also clearly visible as 
  two circular enhancements corresponding to both the exposures.
  Other features of the XMM-OM images, anti-clockwise from top-left:
  (a) loops caused by internal reflection inside the telescope; 
  (b) readout streak for bright sources; 
  (c, e) false detections at the corners;
  (d) cosmic ray bit-flips at different locations 
  in the image. 
}
  \label{fig:af}
\end{figure}

\begin{figure*}
	\hspace*{-0.3cm}\includegraphics[width=0.6\textwidth]{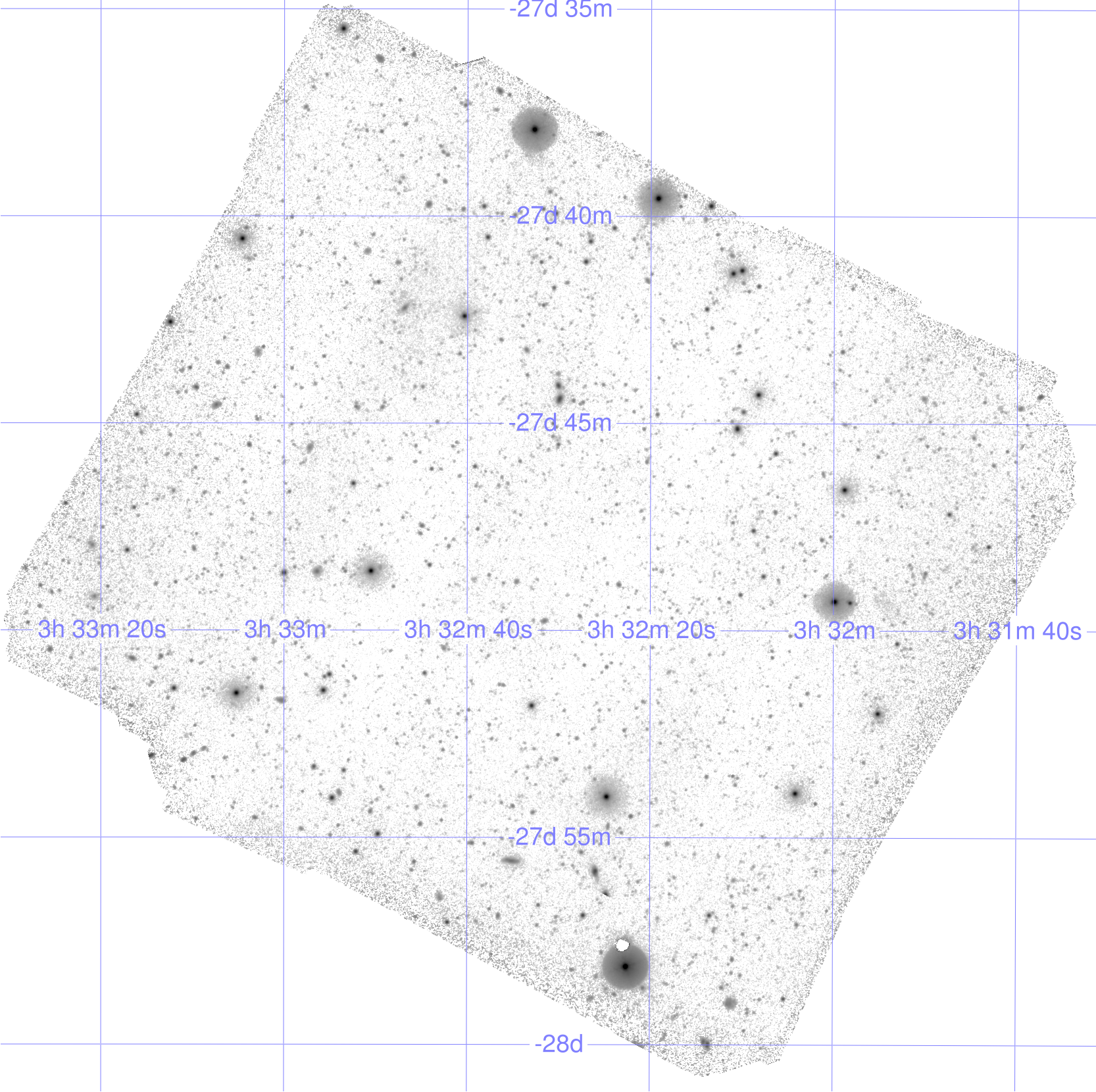}
    \caption{The UVW1 mosaiced image of the CDFS covering a sky-area of 
    $\simeq 395$ arcmin\textsuperscript{2}. It is obtained by co-adding 13 full-frame 
    and 170 rudi-5 exposures of the CDFS.}
    \label{fig:uvw1_img}
\end{figure*}

The XMM-OM data were obtained from the \textit{XMM-Newton} Science
Archive\footnote{\url{https://www.cosmos.esa.int/web/xmm-newton/xsa/}}
(XSA). The reduction and processing were carried out using version 17 of
the standard \textit{XMM-Newton} Science Analysis System
\footnote{\url{https://www.cosmos.esa.int/web/xmm-newton/sas}}
(SAS) tools. 

The SAS script \textsc{omichain} is designed to execute various SAS 
tasks on 
the UVW1 images. These tasks include operations like obtaining the
tracking history of the telescope movement, flat-fielding, 
correcting for modulo-8 (mod-8) noise, source-detection and photometry.
Later into the chain, quality maps are created, the images are projected 
onto the sky plane and aspect correction is done for the source-lists 
and 
the sky-images \citep[see][for details about the tasks making up the chain]{2012MNRAS.426..903P}.

We start with removing the cosmic rays from the raw images before they
are
fed to the \textsc{omichain}. The detector counts from a source are 
saved as a
16-bit number in the instrument memory. A high energy cosmic ray passing
through the
instrument may cause a high bit in the memory to flip and increase the 
count significantly and trigger a false detection 
(Fig. \ref{fig:af}(d)) during
photometry which in the end may add spurious sources to the catalogue. We used code developed 
by \citet{2004PASP..116..148P} to remove most of the cosmic rays. This code leaves
some less significant bit-flip effects in our images, which were removed
manually.

The cosmic-ray-corrected raw images were processed by \textsc{omichain}
up to the mod-8 correction step, after which they were taken out of
the pipeline for additional corrections on top of the standard SAS data
processing. 
First we got rid of the background scattered 
light feature at the center of the image (Fig. \ref{fig:af}).
It was removed by using
a template made using multiple UVW1 images from different fields as explained in \citet{2021MNRAS.506..473P}.
Second, the readout streaks (if any) coming out of the bright sources 
(Fig. \ref{fig:af}(b)) were corrected. The readout streaks are formed by 
photons arriving during 
frame transfer of the CCD \citep{2017MNRAS.466.1061P}. 
These are removed by subtracting the excess count rate of the affected 
columns from the images.
The loopy artefacts shown in Fig. \ref{fig:af}(a), 
come from reflection of bright stars outside the field of view by
a chamfer in the detector window housing inside the telescope
\citep{2001A&A...365L..36M}. These artefacts were removed by masking the 
loops in the corresponding images and applying the same masks for 
individual exposure maps.

After correcting for all the artefacts, the images were then distortion
corrected\footnote{This corrects for the small offsets of the pixel position
from a linear scale \citep{2001A&A...365L..36M}}, aligned with the equatorial coordinate frame of 
COMBO-17 \citep{2004A&A...421..913W, refId4} 
then rotated and re-binned into sky coordinates using the SAS task 
\textsc{omatt}.
A mosaiced UVW1 image was created from individually processed images
by using the SAS task \textsc{ommosaic}. 
The resulting image is shown in Fig. \ref{fig:uvw1_img}.
The edges of the final UVW1 image are the regions of minimum exposure
times and 
hence maximum noise. There is a possibility that noise spikes get detected
as
sources in these regions of low S/N, giving rise to false detections.
To avoid these this, we masked the pixels having an 
exposure time less than 5 ks (which is equivalent to removing a total area of $\simeq 6.43$ arcmin\textsuperscript{2}).
Our corrected image covers a total area $\simeq 395$ arcmin\textsuperscript{2}.
The mosaiced UVW1 image was then fed to the SAS task \textsc{omdetect},
which performs
photometry and source detection, using dedicated algorithms for
identifying point and extended sources. A range of apertures from 2.8 to
5.7 arcsec are used based on brightness of individual sources as well 
as their proximity on the image.
We refer to \citet{2021MNRAS.506..473P} and \citet{2012MNRAS.426..903P} for the details on the source detection. 
We finally end up with a UVW1 source-list containing 3277 sources at $3 \sigma$.

\section{Eddington bias, confusion and Completeness}
\label{sec:3}

\subsection{Eddington bias and flux boosting}
\label{sec:3.1}

There is a likelihood for a source to get confused with another 
source in a densely packed region of the field or it might not be detected
at all due to limitations in the source detection algorithm. 
Therefore, once the final image is produced, it is necessary to
precisely quantify different biases and determine the level of confusion
in the data that are used to construct luminosity functions.

The errors in the photometric flux measurement give rise to the
Eddington bias \citep{1913MNRAS..73..359E}.
Due to measurement errors, more faint magnitude sources are scattered into slightly brighter (or less faint) magnitude bins, compared to sources of slightly brighter magnitudes that are being scattered into fainter
magnitude bins. This causes an 
overestimate in the number of faint sources and biases the 
calculations towards their luminosities.
Precise estimation of this bias is more important for galaxies close
to the detection limits.

\subsection{Source confusion}
\label{sec:3.2}
In a densely packed 
region of the image, two or more sources may be so close that they 
can not be distinguished individually. Such sources could get identified as
a 
single bright source because of their accumulative flux. There is a chance
that 
these false detections get incorrectly identified with counterparts in other
catalogues and end up in the final source-list hampering the final results.
This effect increases with the number of detections and hence sensitivity 
of the instrument. Note for comparison that \citet{2009ApJ...697.1410L} calculated an incompleteness
of 21 per cent due to source confusion in the  \textit{GALEX} NUV filter.

In order to quantify the combined effect of the biases and confusion, we 
calculate the completeness of a sample as a function of magnitude.

\subsection{Completeness simulations}
\label{sec:3.3}

\begin{figure}
  \hspace*{-0.1cm}\includegraphics[width=0.95\columnwidth]{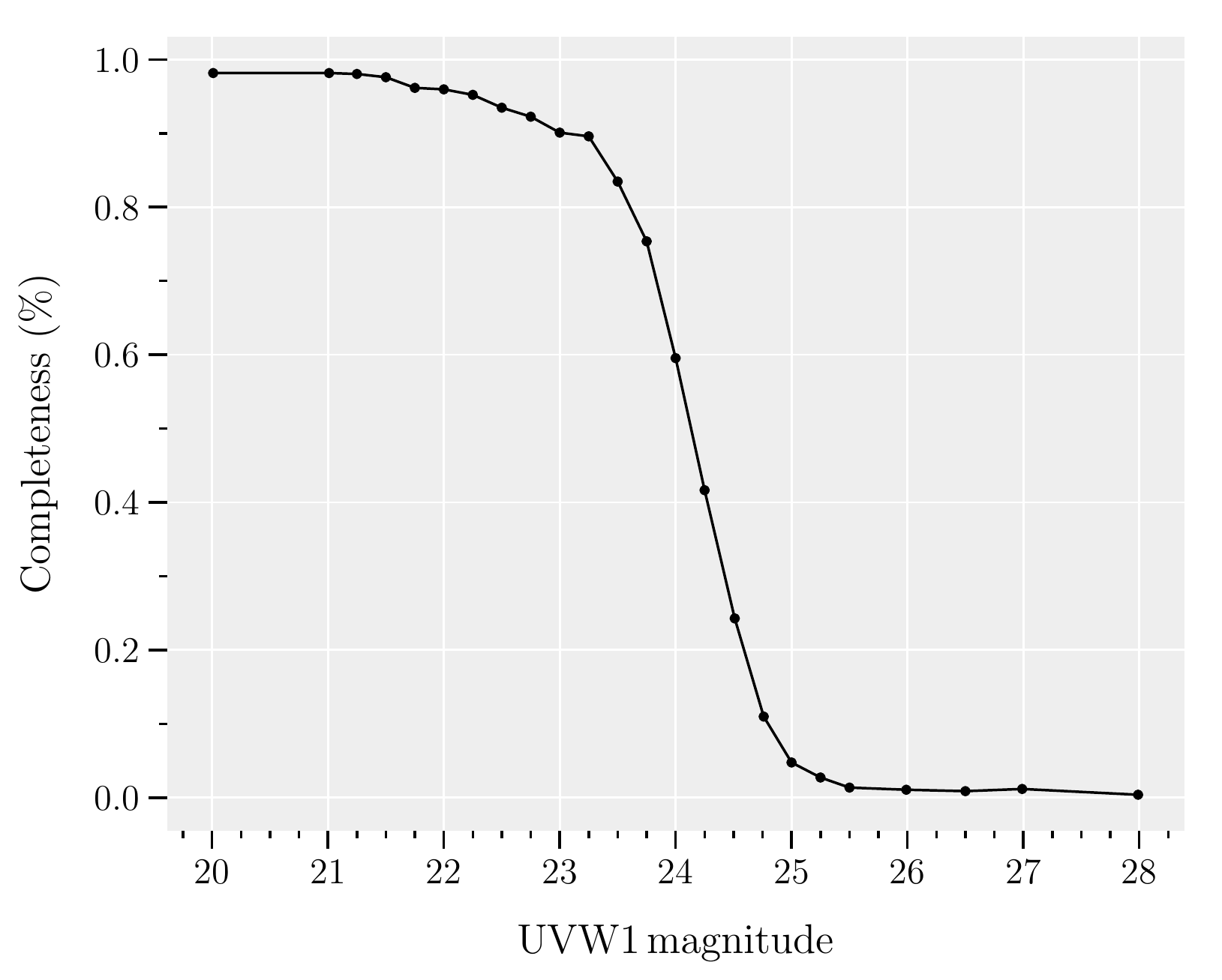}
  \caption{Completeness of the source detection as a function of UVW1
  magnitude, as determined from the simulations described in section 
  \textbf{\ref{sec:3.3}}. 
  The black data points represent the fraction of recovered
  simulated galaxies at each input UVW1 mag.}
  \label{fig:comp}
\end{figure}

We define completeness as - the fraction of sources detected as a function
of the magnitude.
Completeness of a survey goes down with decreasing signal to noise.
For example at the edges of the images, because of the low S/N ratio, 
some sources are missed by the detection algorithm. 

To estimate the completeness of our sample, we use a common technique used
in blank field studies : we simulate artificial galaxies having
properties similar to real galaxies, randomly distribute these 
synthetic galaxies on the image, and detect them using the same
detection process that we use to create the UVW1 source-list.

We simulate 1000 point sources at each input magnitude ranging from
UVW1 mag 21 to 25.5 in steps of 0.25 mag. We extend the range to UVW1 20 mag on the brighter side.
In order to simulate the effect of Eddington bias close to the flux limit 
of the survey, 
and to better assess the source-confusion, we further extend the 
magnitude range to UVW1 mag 28.0 with wider step sizes.  
The sources are inserted 
in the image at random positions one at a time, and if a source is
detected less than 1 arcsec away from the inserted source position, it 
is considered retrieved.

\begin{figure}
  \hspace*{-0.5cm}\includegraphics[width=1.07\columnwidth]{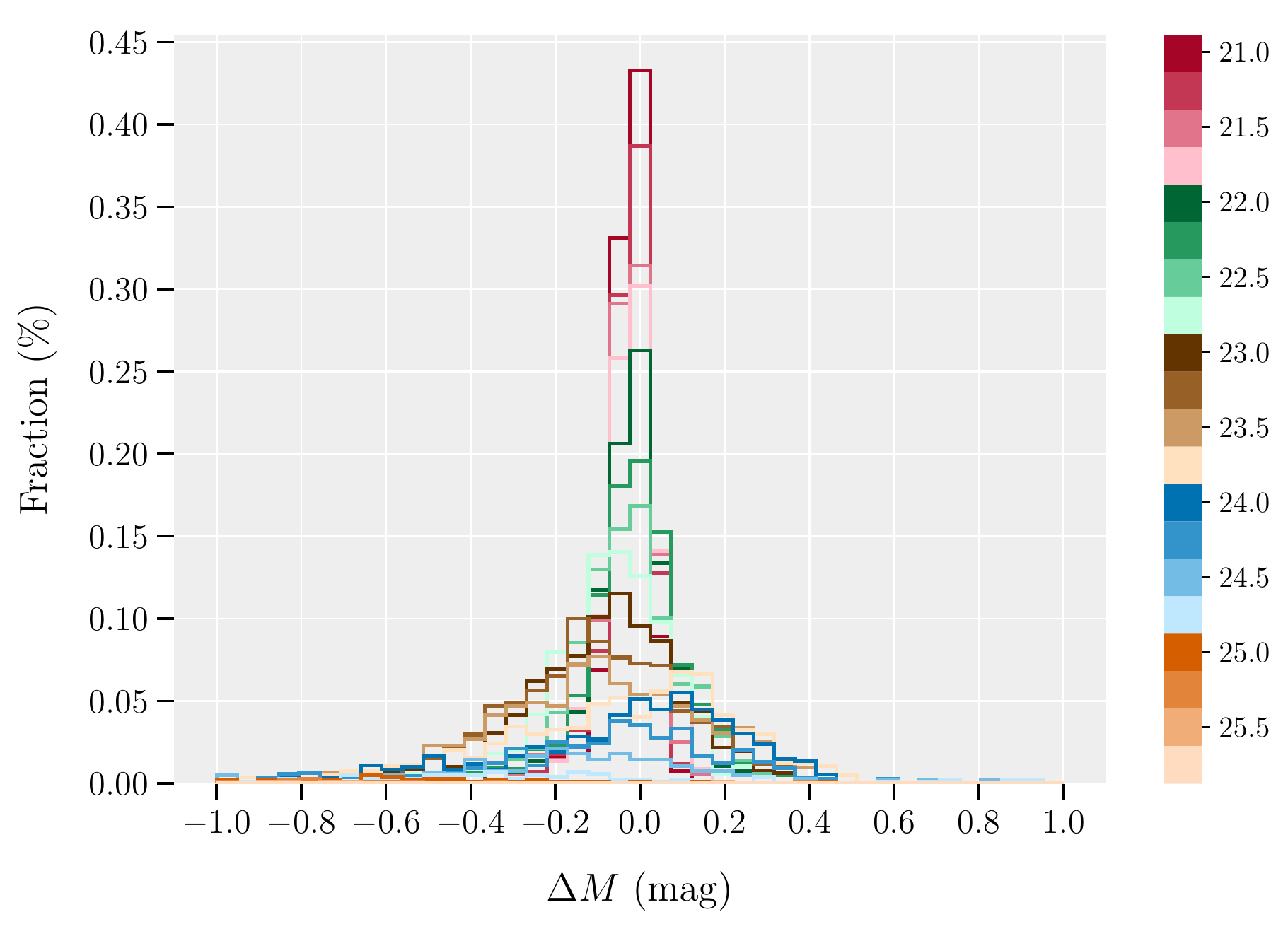}
  \caption{The estimated distribution of the magnitude errors in the 
  detection process at each simulated UVW1 magnitude. Each histogram 
  represents the distribution of offsets of the recovered magnitudes
  from the simulated input magnitudes. The simulated 
  UVW1 input magnitudes are shown by different colours scaled according to
  the color-bar at the right.}
  \label{fig:hist}
\end{figure}

We obtain three pieces of information from the simulations: 
1) The fraction of the injected sources successfully recovered is 
recorded for a given magnitude. We use this fraction as a measure of
\textit{`completeness'}. The fraction of recovered sources 
as a function of UVW1 magnitude is plotted in Fig. \ref{fig:comp}.
The completeness measure is included in our calculations as shown in 
section \ref{sec:6}. 
2) The \textit{`confusion limit'} of the survey is the magnitude beyond
which the source detection process cannot be trusted due
to large source-confusion. We measure the degree of source confusion 
at the magnitude limit of our survey to see if it is confusion limited.
It can be seen from Fig.
\ref{fig:comp}, that the fraction of recovered sources drops steeply 
before the curve asymptotically flattens at the level of 1.5 per cent, which we consider to be the level of source confusion in our image.
This implies that the sensitivity of the detector falls off and the survey 
reaches a sensitivity limit before it becomes limited by confusion.
3) We also obtain an \textit{`error distribution'} from these simulations
which records the distribution of magnitude difference between the injected
and recovered sources. Fig. \ref{fig:hist} shows the error 
histograms at each input magnitude as a function of magnitude offset between
the injected and recovered sources. This
error distribution is later incorporated into our calculations while fitting
the Schechter function to the data using maximum likelihood (for details 
see section \ref{sec:6.2}).
The completeness curve informs our choice of appropriate
survey depth - the faint magnitude limit beyond which no sources were included
in our calculations to produce a binned LF and to estimate the Schechter
function parameters in section \ref{sec:6.1} and \ref{sec:6.2}.

From the completeness measurements, our catalogue is found to be > 97 per 
cent complete for
UVW1 magnitude $\leqslant$ 21.5 and 80 per cent complete at UVW1 magnitude 
$= 23.6$.
The completeness falls 
below 10 per cent at UVW1 magnitude 24.75 and below 5 per cent for UVW1 25 mag (Fig. \ref{fig:comp}). 
Beyond this magnitude
the recovery of sources starts becoming insensitive to the input magnitude.
It can also be seen from Fig. \ref{fig:hist}, that at UVW1 25 mag 
almost all the detections are fainter sources riding positive noise
excursions.
The completeness goes further down to $\simeq 1.5$ per cent for UVW1 magnitudes fainter than 25.5 mag 
(Fig. \ref{fig:comp}). 
We consider 25.5 mag to be the faintest
possible detection limit for our survey. 
As we go fainter, beyond UVW1 25.5 mag, in Fig. \ref{fig:comp}
we are more likely to detect sources because of source
confusion than because of faint real sources on top of noise spikes.

From this analysis we conclude here that our survey becomes insensitive at 
UVW1 25.5 mag, before it becomes confusion limited. In fact with more exposure time on the CDFS, we could have a survey that goes deeper in UVW1 magnitudes.
To make sure we have a very secure UVW1 source-list, make a conservative cut at UVW1 magnitudes 24.5 (one magnitude brighter than the detection limit) as the magnitude limit for our survey, 
where the probability of detecting a source is 15 times the underlying level of source confusion. This magnitude limit 
corresponds to UVW1 mag
24.46 after considering the effect of Galactic foreground extinction explained in section \ref{sec:5.1}.

\section{Cross-correlations to Ancillary data}
\label{sec:4}

\begin{figure}
  \centering
  \hspace*{-0.1cm}\includegraphics [width=0.95\columnwidth]{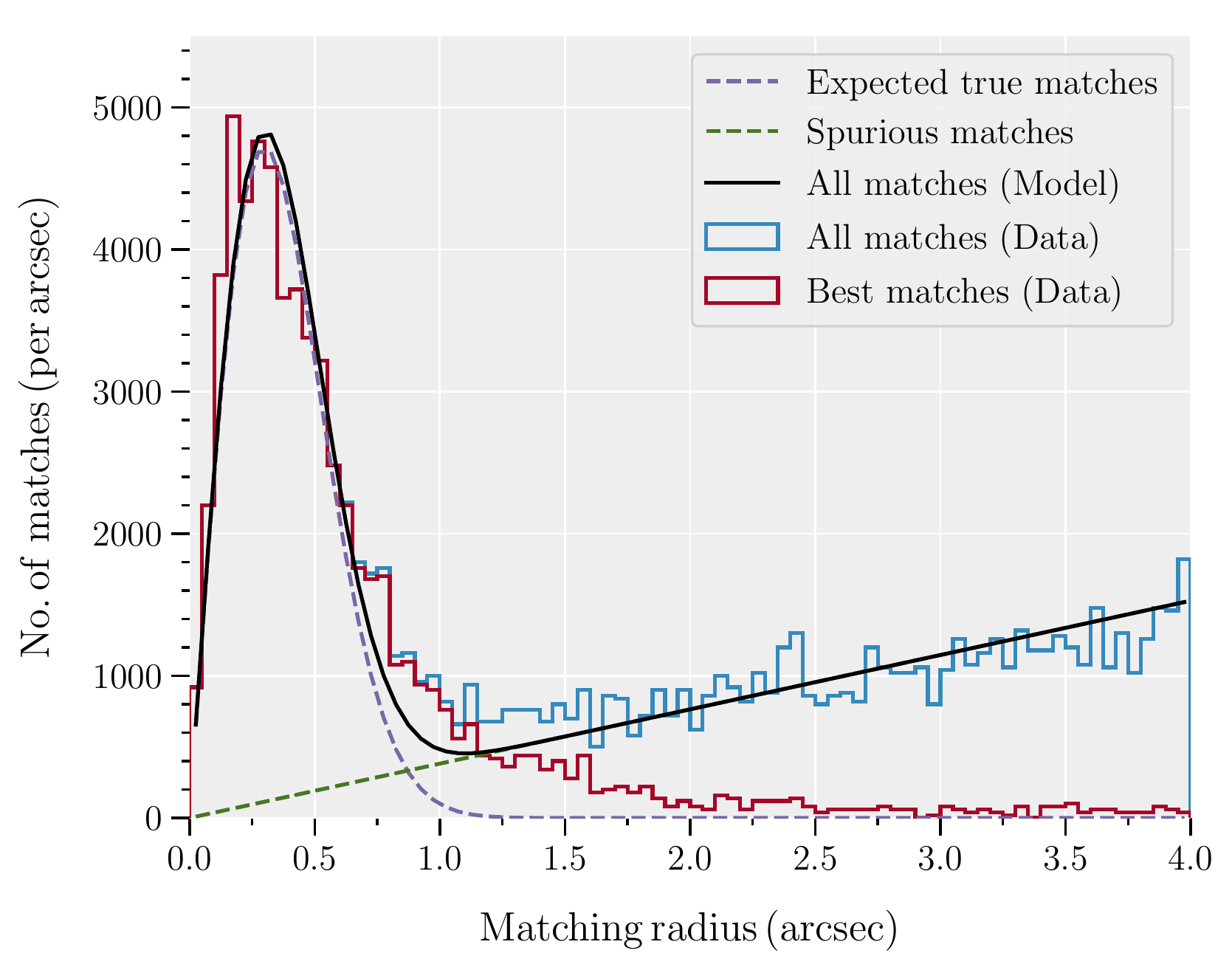}
  \caption{The distribution of angular offsets between all detected
  sources in the UVW1 source-list and their closest counterparts in
  the COMBO-17 catalogue is represented by the red
  histogram.
  The offset distribution of all the matches found in
  COMBO-17 for each source in the UVW1 source-list is 
  represented by blue histogram. The bin size of 0.05 arcsec is used 
  for both histograms.
  The black solid line is the expected model obtained by simultaneously 
  fitting a Rayleigh distribution (dotted purple curve) and a line 
  (dotted green curve) to the distribution of offsets w.r.t. all matches.}
  \label{fig:spurious}
\end{figure}

The UVW1 source-list is cross-correlated with various catalogues 
to acquire some additional information on the sources, primarily
redshifts. 
An appropriate matching radius for the cross-correlation is needed 
for that 
purpose or else a very low matching radius may produce an inadequate 
number of matches and 
a large matching radius may give rise to unwanted spurious matches with
other
catalogues.
We first determine an appropriate matching radius by comparing the expected
and 
observed distribution of cross-correlated sources. The ancillary data from 
other catalogues is summarised next.
We then conclude this section by using the ancillary data
to assign redshifts to our 
sources, and also identify and remove any bright AGN or stars present in the
source-list.

\subsection{Spurious cross-correlations}
\label{sec:4.1}

We match our UVW1 source-list to the COMBO-17 \citep{2004A&A...421..913W,refId4} catalogue, with a variable angular offset up to 5 arcsec.
The distributions of angular offsets corresponding to a best match (closest 
counterpart within the offset limit) and all matches (all counterparts within 
the offset limit) are plotted in Fig. \ref{fig:spurious}. 
These distributions are plotted in blue and red solid lines respectively, with
the x-axis truncated to 4 arcsec offsets.

A Rayleigh distribution is predicted for the probability distribution 
of the angular offsets, given that the uncertainties in their positions 
have a Gaussian distribution \citep{2012MNRAS.426..903P}. The number of spurious matches (false 
counterparts) should grow linearly as a function of matching radius.
The total expected distribution of matches (actual as well as spurious)
is obtained by combining the Rayleigh distribution corresponding to the 
actual sources and a straight line corresponding to unrelated matches.
This distribution as a function of offsets ($x$) should take the form,
\begin{equation}
  D(x) = \,
  A\, \frac{x}{\sigma^2} \,
  \mathrm{exp}\left({-\frac{x^2}{2\sigma^2}}\right) + m\,x.
\end{equation}
This predicted distribution of all offsets is fitted to the distribution 
of all matches, with $A,\,\sigma,\,$ and $m$ as free parameters.
The fitted model is represented by the solid black line in Fig. 
\ref{fig:spurious}.
The dashed purple curve is the Rayleigh distribution fit 
and the straight line is plotted in dashed green.
The fit parameters values are $2330 \pm 60$ sources, 
$0.299 \pm 0.006$ arcsec and $382.2 \pm 14.6$ sources per square arcsec 
for $A,\,\sigma,\,$ and $m$ respectively.

It is clear from Fig. \ref{fig:spurious} that, 
the modelled distribution of true matches
drops asymptotically after 1 arcsec matching radius, and the ratio of 
number of spurious to true matches grows rapidly after that.
So, 1 arcsec seems to be reasonable choice for a cross-matching offset radius.

In order to assess the quality of the cross-correlation for our best 
matches
and to gauge the likelihood that 
the matched sources happen to be within the error radius of the UVW1 
sources 
just by chance, we use Monte-Carlo simulations.
We simulate a random sample of linearly distributed
offsets up to a maximum offset of 1 arcsec from the fitted straight 
line.
Another random sample of potential matches is simulated from the 
true 
(Rayleigh) source
distribution of all matches within the offset of 1 arcsec.
These samples are compared to each other and we note 
the number of spurious offsets, smaller than the expected true offsets, 
and 
consider only this number of actual spurious sources may 
make their way into our source-list. 
Using these simulations we find that $77 \pm 13$ ($3$ per cent) 
out of total 2544 matches, will really be spurious.
The errors are 95 per cent confidence intervals 
obtained through bootstrap re-sampling. So adopting 1 arcsec as the offset
radius will give rise to $3$ per cent of the total matched 
sources being spurious. 
An important remark needs to be added here : we have ignored clustering of 
faint sources.
Since we have compared our catalogue with a deeper ground based survey, the 
background source distribution will be dominated by sources
that are fainter than the UVW1 detected sources. For the case when there is a closer, 
but fainter background source than the correct counterpart, within an arcsec, it is quite 
likely that the fainter source will not even be detected and the correct 
counterpart will be chosen. So, the actual fraction of sources that are matched spuriously will be smaller than $3$ per cent.
We think it is a reasonable compromise, and adopt 1 arcsec as the offset 
radius 
when we cross-correlate our catalogue with other catalogues unless stated 
otherwise.

\subsection{Ancillary data}
\label{sec:4.2}

Our final image extends outside the CDFS as defined by the Chandra X-ray survey, so our ancillary data also includes
surveys targeting the Extended-CDFS \citep[E-CDFS][]{2005ApJS..161...21L} in addition to those
exclusively from CDFS. 
The E-CDFS has been covered by almost 50 bands from UV to mid-infrared
(MIR) (see Table 1 and 2 in \citet{2014ApJ...796...60H}) in different surveys.

We will briefly summarise the catalogue and surveys used in this work.
\citet{2014ApJ...796...60H} used photometry from UV to MIR to estimate the
photometric redshifts in the E-CDFS. We use their catalogue for  spectroscopic and photometric redshifts.
We also use the VLT\footnote{the details can be found at
\url{http://www.eso.org/sci/activities/garching/projects/goods/MasterSpectroscopy.html}} 
master catalogue of spectroscopic redshifts compiled from various 
spectroscopic campaigns in the CDFS up until 2012-13. \citep{refId1,2004ApJS..155..271S,refId6,refId7,refId2,refId5,Silverman_2010,refId7}. 
FIREWORKS \citep{2008ApJ...689..653W} is a $K-$band selected
catalogue for the CDFS, containing photometry in several bands. 
It provides both photometric and spectroscopic redshifts.
The catalogue provided by \citet{2011ApJ...742....3R} is the source 
of most of the photometric redshift used in this work. This catalogue
is constructed using 7 different (see Appendix A.1 in 
\citet{2011ApJ...742....3R}) catalogues in the E-CDFS to 
derive the photometric redshifts.
The COMBO-17 survey imaged CDFS with the Wide Field Imager at the MPG/ESO
2.2m telescope, in 17 passbands to obtain photometric redshifts for galaxies identified at $R_{\mathrm{Vega}} < 24$. 
\citep{2004A&A...421..913W, refId4}.
The MUSYC survey, which used photometry in 32 bands to estimate redshifts in the E-CDFS \citep{2010ApJS..189..270C,2009ApJS..183..295T} also contributes 
a few photometric redshifts to our catalogue.
The Gaia DR2 \citep{2018A&A...616A...1G} is the second data release catalogue 
from observations of the Gaia observatory. We use it to identify the stars in our source-list.

\subsection{Cross-correlation to optical and mid-infrared counterparts}
\label{sec:4.3}
Here we match our source-list with the ancillary data listed in 
the 
previous sub-section, using the 1 arcsec radius deduced in sub-section 
\ref{sec:4.1}.
We start with removing the stars and then obtain the redshifts for our sources. We remove bright AGN present in the catalogue at the end.

\subsubsection{Stars}
The presence of stars in the source-list is irrelevant to the 
extra-galactic UVLF. 
The CDFS is known for its isolated position towards the Galactic halo, 
where 
the stellar population is comparatively aged, so the UVW1 sample will have little contamination from Galactic stars.
Nevertheless we cross correlated our catalogue with
the Gaia DR2 \citep{2018A&A...616A...1G} to identify the stars by matching it with our
source-list and removing sources with significant proper motions.
COMBO-17 data were also 
employed to remove any stars from the source-list.
They used stellar templates from different libraries to 
identify the stars. 
We removed 155 sources identified as stars in total, 
leaving 3122 sources.

\subsubsection{Redshifts} 

Once the stars were removed we cross-matched the remaining sources with the
redshift catalogues.
Most of our spectroscopic redshifts are from the \citet{2014ApJ...796...60H} catalogue. We obtain more than 1400 spectroscopic redshifts from their catalogue.
The VLT master catalogue provides 212 spectroscopic redshifts. 
The FIREWORKS catalogue is used only for its spectroscopic redshifts.
Appropriate quality flags provided by each catalogue have been used to
apply conservative quality cuts (see Table \ref{tab:redshift}), in order to make sure we get the most accurate redshifts. 
Regarding the photometric redshifts, \citet{2011ApJ...742....3R} 
contributes 838 photometric redshifts. 
This is followed by catalogues 
from COMBO-17 and MUSYC surveys providing 68 and 28 redshifts, respectively.
We used the \citet{2014ApJ...796...60H} catalogue again to acquire 66 photometric 
redshifts in addition to the spectroscopic ones.
We applied a quality cut to only allow photometric redshifts for which 
the 68 per cent confidence region around the peak redshift has a width < 0.2.
The list of all the catalogues used for redshifts with references, 
number of
contributed redshifts and the quality cuts can be found in Table
\ref{tab:redshift}.

\begin{figure}
  \hspace*{-0.2cm}\centering
  \includegraphics [width=0.95\columnwidth]{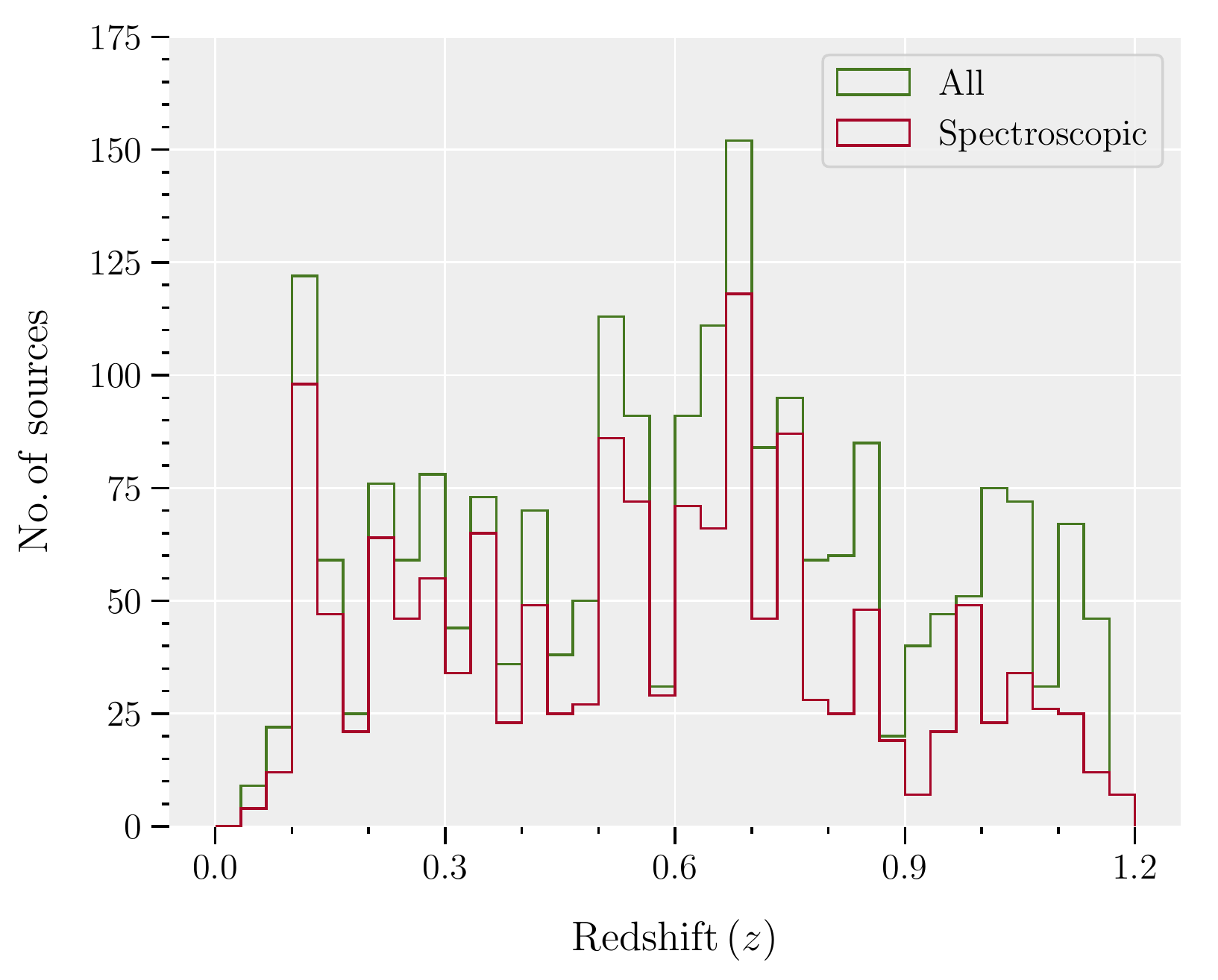}
  \caption{Redshift distribution of the CDFS sample. The 
  spectroscopic redshifts are represented in red colour and total number is shown in green.}
  \label{fig:z_distro}
\end{figure}

\begin{table}
  \centering
  \caption{Catalogues used for spectroscopic and photometric 
  redshifts, along with the number of redshifts and quality flags (QF) 
  used 
  for each catalogue. \texttt{z68} represents the 68 percent confidence interval around each photometric redshift in the photometric catalogues.}
  \label{tab:redshift}
  \begin{tabular}{llr}
    \hline\hline
    \noalign{\vskip 0.5mm}
    Source Catalogue &
    Number$^a$ &
    QF \\
    \hline
    \noalign{\vskip 0.5mm}
    \multicolumn{1}{@{}l}{Spectroscopic redshifts}\\
    \citet{2014ApJ...796...60H}         & 1403  & 0,1     \\
    \citet{refId1}         & 30    & 3,4     \\
    \citet{2004ApJS..155..271S}          & 38    & 2       \\
    \citet{refId6}              & 21    & 1       \\
    \citet{refId7}            & 34    & 2       \\
    \citet{refId2}                      & 17    & A       \\
    \citet{refId5}           & 63    & A       \\
    \citet{Silverman_2010}       & 9     & 2       \\
    \citet{refId8}               & 6     & 1       \\
    \citet{2008ApJ...689..653W}         & 3     & 1       \\
    \hline
    \noalign{\vskip 0.5mm}
    \multicolumn{1}{@{}l}{Photometric redshifts}\\
    \citet{2011ApJ...742....3R}            & 832   & \texttt{z68} $<0.2$       \\ 
    \citet{2014ApJ...796...60H}            & 66    & \texttt{z68} $<0.2$       \\
    \citet{refId4}          & 46    & \texttt{z68} $<0.2$,      \\
    & & $R < 24$ mag    \\ 
    \cite{2010ApJS..189..270C} & 17    & \texttt{z68} $<0.2$,    \\
    & & \texttt{QF} $\leq 1$       \\
    \cite{2009ApJS..183..295T}             & 13    & \texttt{z68} $<0.2$       \\
    \cite{2015ApJ...801...97S}           & 1     & \texttt{z68} $<0.2$,    \\
    & & \texttt{QF} $= 0$       \\

    \hline
  \end{tabular}\\
  \begin{minipage}{0.86\columnwidth}
      \textsuperscript{$a$}{Represents the numbers before the bright
      AGN are removed.}
  \end{minipage}
\end{table}

\subsubsection{Active Galactic Nuclei} 

The accretion discs around the central super-massive black holes in
active galaxies are known to emit large quantities of UV radiation, which dominates
the UV flux coming from star formation in these galaxies.
The inclusion of these galaxies with UV bright active galactic nuclei
(AGN) in the sample can cause erroneous estimation of the UV LF. 
These sources affect the shape of the UV LF especially at the bright 
end because very bright galaxies are extremely rare.
The CDFS has a significant number of bright AGN in the redshift range
of our interest, that can contaminate the UVW1 sample.
Assuming that AGN bright in UV fluxes will also have X-ray 
emission,
the X-ray catalogues from \citet{2016ApJS..224...15X} and \citet{2017ApJS..228....2L}
are used to identify the AGN present in our sample.
Any bright AGN
making their way into the sample were removed by using a luminosity
cut-off $10^{42}$ ergs s$^{-1}$ in the full x-ray band (0.5 - 8 KeV).
We make detailed checks on the effect of AGN contamination on our 
sample
in the redshift range we explore, in Appendix~\ref{sec:agn}.

After removing the potential AGNs, a sample of 2518 galaxies is selected with redshifts of the highest quality, 1559 of which are spectroscopic.
The redshift distribution of our final galaxy sub-sample truncated to
redshift 1.2 is shown in Fig. \ref{fig:z_distro}.
A section of the source-list is presented in Table \ref{tab:cat} and
the full table of sources ($z = 0.6 - 1.2$) in machine readable form is available online.

\begin{table}
  \centering
  \caption{The UVW1 source list used for this paper. The positions and 
  apparent UVW1 AB magnitudes are calculated using \texttt{omdetect} 
  (see section \ref{sec:2}), the redshifts are obtained by cross-matching
  with other catalogues (see section \ref{sec:4} and Table \ref{tab:redshift}). The full table is available in the machine readable form with the online version of the paper.}
  \label{tab:cat}
  \begin{tabular}{cccr}
    \hline\hline
    \noalign{\vskip 0.5mm}
    RA (J2000)  &
    DEC (J2000) &
    z    &
    UVW1 mag \\
    \multicolumn{1}{c}{deg} &
    \multicolumn{1}{c}{deg} &
    &
    \\
    \hline
    \noalign{\vskip 0.5mm}
    53.2170 & -27.7417 &  0.80   & 21.42  \\
    53.0706 & -27.6580 &  1.00   & 21.76  \\
    53.2845 & -27.8815 &  0.74   & 21.91  \\
    53.0778 & -27.6676 &  0.61   & 21.93  \\
    53.1258 & -27.8849 &  0.64   & 22.00  \\   
    \hline
  \end{tabular}\\
\end{table}

\section{Corrections}
\label{sec:5}
\subsection{Galactic foreground extinction}
\label{sec:5.1}

The radiation coming from extra-galactic objects is absorbed and/or 
scattered by 
the interstellar dust in the Milky way. As a result, the incoming light is 
extincted by a certain magnitude. 
Galactic extinction for a given photometric band is given by,
$A_{\lambda} = k\left( \lambda \right) \, E \left(B - V\right)$, where 
$k\left( \lambda \right)$ is the extinction coefficient determined by 
the Galactic extinction curve \citep{1989ApJ...345..245C} and 
$E \left(B - V\right)$ is determined from the dust map of the sky. 

We use an all-sky dust extinction map from \citet{1998ApJ...500..525S} to
obtain 
$E \left(B - V\right)$, and Table 6 from \citet{2011ApJ...737..103S}
to calibrate the above relation.
The correction for dust-extinction in the UVW1 band,
$A_{\mathrm{UVW1}}$ is applied in equation \ref{eqn:absmag}
after applying the K-correction (section \ref{sec:5.2}).
Its value along the direction of CDFS is estimated to be 0.039.
This low value of extinction is supported by a low value of the 
Galactic HI column density of around 8.8 $\times$
10\textsuperscript{19} cm\textsuperscript{-2} \citep{1992ApJS...79...77S}.

\subsection{K correction}
\label{sec:5.2}
The observed spectral energy distributions (SEDs) appear different from 
the rest-frame SEDs 
as they are red-shifted due to the expansion of the Universe.
So, the measurements taken in a particular waveband sample different 
parts of the 
SED of a galaxy, depending upon the redshift of that galaxy.
Thus, when we
compare galaxies in a given observed waveband 
(say $\left[\nu_{1}, \nu_{2}\right]$) at different redshifts, 
we are actually looking at two different rest-frame wavebands. 
This effect is most severe for very distant galaxies.

To correct for this offset, we use an additive term
called K-correction,
while calculating the absolute magnitudes of the galaxies from apparent magnitudes. 
The method used to derive these K-corrections as a function of redshift
from the appropriate best-fit templates is explained in detail 
in \citet{2021MNRAS.506..473P}.
Once the K-corrections $K\left(z\right)$ are calculated, 
the following relation is used to calculate the absolute 
magnitude for each source
\begin{equation}
  M_{1500}\left(z\right) = m-5\log{\left(\frac{d_L\left(z\right)}
    {\mathrm{Mpc}}\right)} -25 -K\left(z\right) - A_{\mathrm{UVW1}},
    \label{eqn:absmag}
\end{equation}
where $d_L$ is the luminosity distance.

\section{Luminosity function}
\label{sec:6}

\begin{figure}
  \hspace*{-0.1cm}\includegraphics[width=0.95\columnwidth]{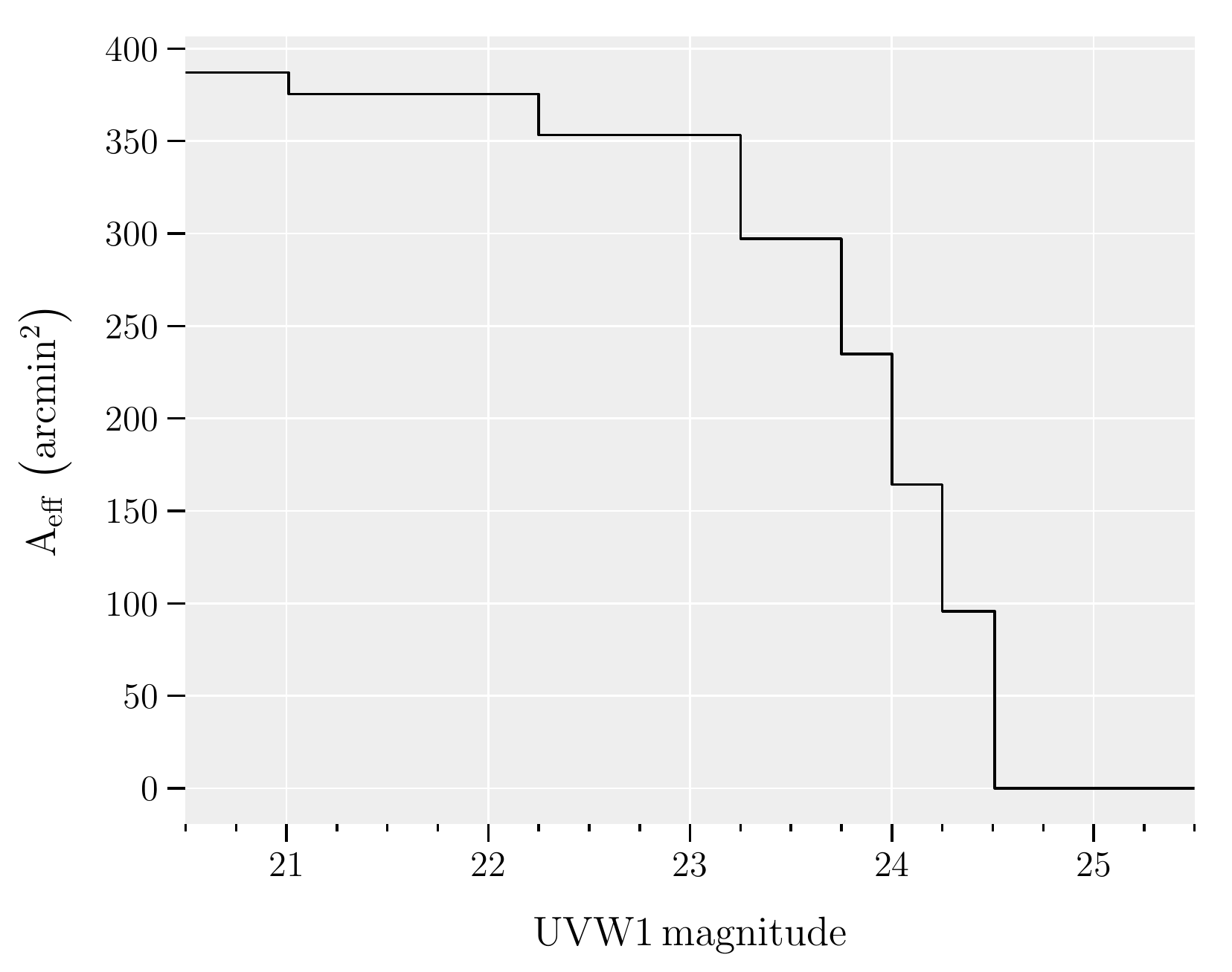}
  \caption{Effective area as a function of UVW1
  magnitude. The completeness of the survey is taken into account using this
  function in the construction of binned luminosity functions 
  as described in section \ref{sec:6.1}.}
  \label{fig:eff_area}
\end{figure}

\subsection{Binned luminosity function}
\label{sec:6.1}

We use the \citet{2000MNRAS.311..433P} method to calculate the binned 
luminosity function.
The number of galaxies $N$, inside a bin of the volume-magnitude 
space bound by redshift interval $z_\mathrm{min}<z<z_\mathrm{max}$ 
and absolute magnitude interval $M_\mathrm{min}<M<M_\mathrm{max}$, is
related to the differential luminosity function $\phi\left(M,z\right)$ 
as
\begin{equation}
  N = \phi\left(M,z\right) 
  \int_{M_\mathrm{min}}^{M_\mathrm{max}}\,
  \int_{z_{\mathrm{min}}}^{z_{\mathrm{max}}}
  \frac{\mathrm{d}V\left(z\right)}{\mathrm{d}z}\, 
  \mathrm{d}z\, \mathrm{d}M,
\end{equation}
assuming that the variations of $\phi\left(M,z\right)$ over redshift and absolute magnitude intervals are small enough to be ignored \citep{2000MNRAS.311..433P}.
The integral on the right hand side of the above equation can be considered as the effective 4-volume $V_{\mathrm{bin}}$ of the volume-magnitude bin. 

We divide the total survey area $A$ into $j$ sub-fields, and add the contribution of 
each sub-field towards the effective volume of the volume-magnitude
bin,
\begin{equation}
  V_{\mathrm{bin}} = 
  \int_{M_\mathrm{min}}^{M_\mathrm{max}}\,
  \sum_{j}\, \int_{z_{\mathrm{min},\, j}}^{z_{\mathrm{max},\, j}}
  \frac{\mathrm{d}V\left(z\right)}{\mathrm{d}z}\, 
  \mathrm{d}z\, \mathrm{d}M, 
  \label{eqn:iv}
\end{equation}
and obtain, $\mathrm{d}V\left(z\right)/\mathrm{d}z$ for each 
sub-field by multiplying its effective area with the
differential comoving volume element,
\begin{equation}
  \frac{\mathrm{d}V\left(z\right)}{\mathrm{d}z} = 
  A_{\mathrm{eff},\, j}\left(M\right) \,
  \frac{c\,H_{0}^{-1}\,d_{L}^{2}}
  {(1+z)^2\,[\Omega_{\lambda} + \Omega_{m} (1+z)^3]^{1/2}}
\end{equation}
where, $A_{\mathrm{eff}}$ is the effective area of each sub-field,
\begin{equation}
  A_{\mathrm{eff},\, j}(m) = 
  A\, C_{j}\left(m\right)\, 
  \left(\frac{\pi}{180}\right)^2.
\end{equation}
Here $A$ is in $\mathrm{deg}^2$ and $C_{j} \left(m\right)$ is the 
completeness fraction for each sub-field.
The total effective area $A_{\mathrm{eff}} \left(m\right)$, plotted in 
Fig. \ref{fig:eff_area}, is a 
step function of apparent magnitude where the steps are the 
the effective areas for the individual sub-fields.

The number of sources $N_{\mathrm{bin}}$ are counted in each bin and divided by $V_{\mathrm{bin}}$ to estimate the binned LF,
\begin{equation}
  \phi = 
  \frac{N_\mathrm{bin}}{V_\mathrm{bin}}.
\end{equation}

\begin{figure*}
    \centering
    \hspace*{-0.2cm}\includegraphics[width=0.87\textwidth]{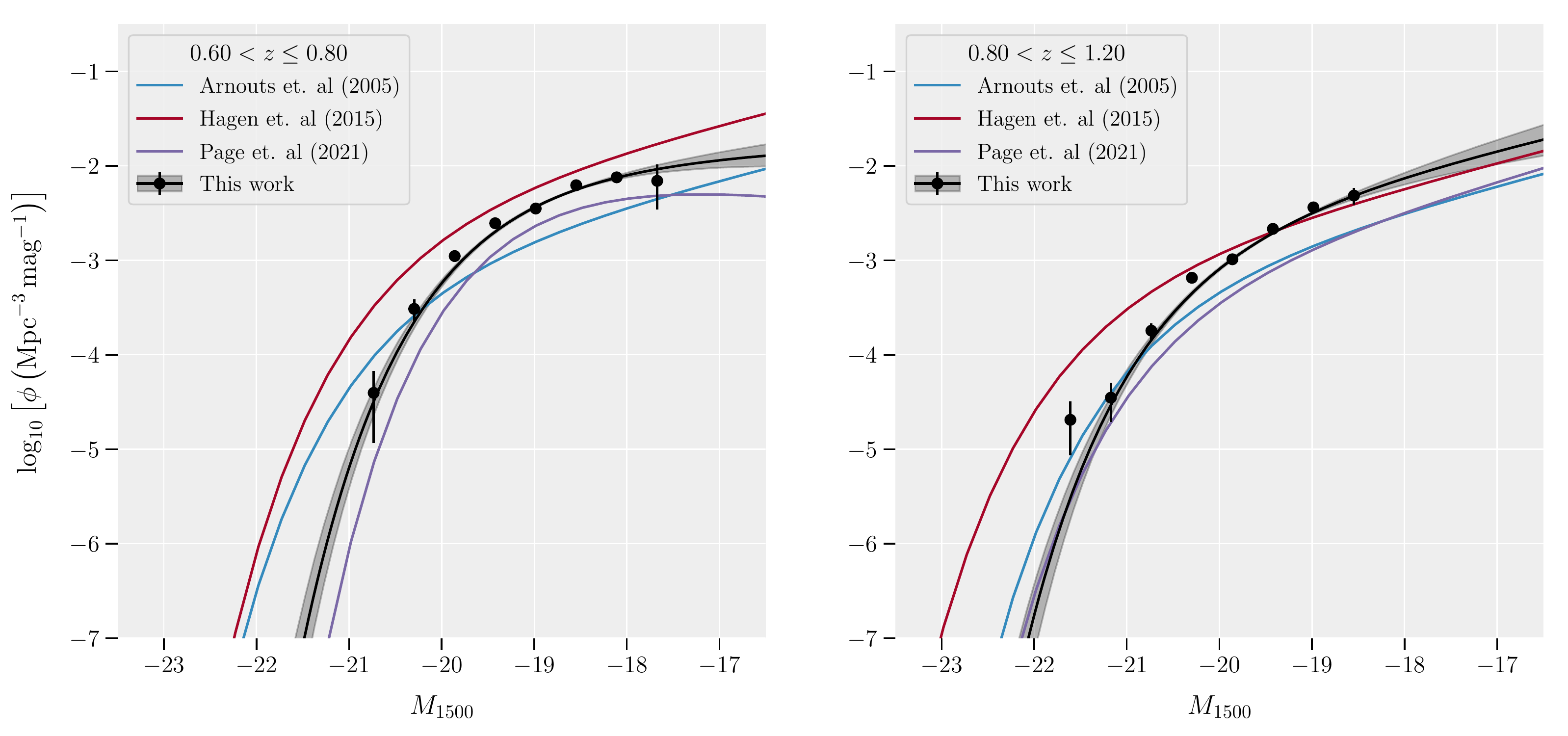}
    \caption{UV luminosity function of galaxies in the redshift
    intervals $0.6 \leq z < 0.8$ in the left panel and $0.8 \leq z < 1.2$
    in the right panel as a function of the 1500 {\AA}  magnitude.
    The data points show the binned number densities measured 
    using the \citet{2000MNRAS.311..433P} method. The black solid
    line is our best-fitting Schechter function 
    derived from the CDFS field as described in Section \ref{sec:6.2}.
    We obtain this curve from the median value of the posterior distribution
    of Schecter function parameters.
    The grey shaded area around the best-fit Schecter function
    represents the $1 \sigma$ (68.26 per cent) uncertainties.
    The blue and red and purple solid lines are the Schecter functions 
    obtained by \citet{2005ApJ...619L..43A}, \citet{2015ApJ...808..178H}
    and \citet{2021MNRAS.506..473P}.}
    \label{fig:lf68}
\end{figure*}

From Poisson's statistics \citep{1986ApJ...303..336G}, we calculate
the uncertainty for $N$ objects and hence the statistical uncertainty in the LF for each bin in the redshift-magnitude space.
Because of the nature of our survey, it may also be subjected to cosmic variance, which we discuss in detail in section \ref{sec:7}.
The resulting luminosity function $\phi$ with units of $\mathrm{Mpc}^{-3}\mathrm{mag}^{-1}$, are shown in Fig. \ref{fig:lf68}

\subsection{Schechter function parameters}
\label{sec:6.2}

To recover more information from a magnitude limited sample of
galaxies, we analyse the galaxy distribution in the redshift-magnitude 
space by comparing it to a galaxy LF model using a maximum 
likelihood estimator. We adopt the Schechter function \citep{1976ApJ...203..297S} 
to model the galaxy LF in each redshift bin. It can be parametrised as a 
function of magnitude $M$ with parameters $\alpha,\phi^{*}$ and $M^{*}$,
\begin{equation}
    \phi(M) \equiv \phi(M;\alpha,\phi^{*},M^{*}) = 
    k\,\phi^{*} \,
    \frac{e^{k(1+\alpha)(M^{*}-M)}}
    {e^{{e^{k(M^{*}-M)}}}},
    \label{eqn:phi}
\end{equation}
where $k = \frac{2}{5}\, \ln{10}$.
The functional form at faint magnitudes is characterised by a power law slope $\alpha$, also called the faint-end slope. 
The power law form 
cuts off at a characteristic magnitude, $M^{*}$, and an exponential 
behaviour follows to brighter magnitudes. $\phi^{*}$ is the 
normalisation.
We use the maximum likelihood approach to fit the above model to our
observations. Our likelihood function for all $N_G$ sources is defined as
\begin{equation}
  \mathcal{L} = \prod_{i}^{N_{G}}p\left(M_i, z_i\right)
\end{equation}
where $p\left(M_i, z_i\right)$ is the probability density for a galaxy $i$,
to be observed with an absolute magnitude $M_i$ in the survey magnitude limits
$[M_{min},M_{max}]$ 
at the object's redshift $z_i$ in the interval $[z_{min}, z_{max}]$.
It is given by
\begin{equation}
  p\left(M_{i},z_{i}\right) = 
  \frac{\phi\left(M_{i}, z_{i}\right)}
  {\int_{M_{\mathrm{min}}}^{M_{\mathrm{max}}}
  \int_{z_\mathrm{min}}^{z_\mathrm{max}}
  \phi\left(M, z\right)\, 
  \frac{\mathrm{d}V \left(z\right)}
  {\mathrm{d}z}\,\mathrm{d}z\, \mathrm{d}M}.
  \label{eq:p}
\end{equation}
The Schechter function parameters can be found by maximising 
$\mathcal{L}$, which is equivalent (and more numerically convenient) to minimising 
the negative logarithm of the likelihood,
\begin{equation}
  S = -2\ln\mathcal{L} = 
  -2 \sum_{i=1}^{N_\mathrm{G}}\ln p\left(M_i, z_i\right).
  \label{eqn:L}
\end{equation}
It is important to note here that the normalisation $\phi^{*}$ can not be
fitted jointly with
$M^{*}$ and $\alpha$, using this framework as it gets cancelled out in equation \ref{eq:p}, and the final likelihood function is independent 
of it. It is determined separately by
asserting that the predicted and observed number of sources is equal.

The measurement errors associated with the data can seriously hamper the
outcomes of the fitting process, if not properly accounted for. The 
effects of these errors may become more severe particularly for our 
Schechter
function model which increases exponentially at bright magnitudes. 
To take into account the uncertainties, we use the error distribution
(Fig. \ref{fig:hist}) derived from the completeness simulations in section
\ref{sec:3.3}. 
The observed galaxy LF
for a population of galaxies can be obtained from the underlying LF
via this error distribution.
We use the formalism discussed in \citet{2021MNRAS.506..473P} to incorporate the 
photometric uncertainties into our calculations. A brief description of
the method is given below.

If a galaxy of true absolute magnitude $M$ is observed with an
absolute magnitude in the range 
$[M^\prime,\, M^\prime+\mathrm{d}M^\prime]$ with
a probability $P\left(M^\prime|M\right)\mathrm{d}M^\prime$,
an underlying LF $\phi\left(M\right)$ should be observed as a 
differential LF 
$P\left(M^\prime|M\right)\phi\left(M\right)\mathrm{d}M$.
In other words, we marginalize our model
over the error distribution of magnitudes $P\left(M^\prime|M\right)$, 
instead of calculating it for single galaxy magnitudes.
This can be included in the maximum likelihood method by modifying 
equation \ref{eq:p}, which now becomes
\begin{multline}
  p\left(M^\prime_{i},z_{i}\right) = 
  \frac{\int
  \phi\left(M_i, z_i\right) P\left(M^\prime|M\right)\mathrm{d}M}
  {\int \int \int \phi\left(M, z\right) P\left(M^\prime|M\right)\mathrm{d}M\,
  \frac{\mathrm{d}V \left(z\right)}
  {\mathrm{d}z}\,\mathrm{d}z\, \mathrm{d}M^\prime}.
  \label{eq:p2}
\end{multline}
We incorporate the completeness of the survey into our likelihood 
formalism by normalising these histograms
with the number of injected sources in completeness simulations. 
We refer to the study by \citet{2021MNRAS.506..473P} for more details of the framework. 
This framework is incorporated in our work by using a Markov Chain Monte Carlo (MCMC) prescription (described in the next paragraph), whereas \citet{2021MNRAS.506..473P} use a different approach.

We minimise equation \ref{eqn:L} and then using MCMC, we obtain the posterior probability distribution for the Schechter 
function parameters around the minimum value of the likelihood. We use
\textsc{emcee} \citep{2013PASP..125..306F} - 
a popular python implementation of the Affine Invarient MCMC Ensemble Sampler \citep{2010CAMCS...5...65G} to apply the MCMC.
In this scheme, a likelihood function as in equation \ref{eqn:L} can be
obtained from a posterior probability distribution function given by,
\begin{equation}
    p\left(\theta\,|\,M^\prime_{i},z_{i}\right)\propto
    p\left(\theta \right)\times p\left(M^\prime_{i},z_{i}\,|\,\theta\right),
    \label{eqn:p}
\end{equation}
where, $\theta = \left(\phi^{*},M^{*},\alpha\right)$ are the model
parameters, and $M^\prime_{i}$ and $z_{i}$ are the the magnitude
and redshift of the galaxies lying inside the redshift bin $z_\mathrm{min}<z<z_\mathrm{max}$.
A uniform uninformative prior distribution is assumed
$p\left(\theta \right)$ for our parameters and the likelihood function 
is given by, $\mathcal{L}= p(M^\prime_i, z_i|\theta)$.

\begin{figure}
  \hspace*{-0.2cm}\centering
  \includegraphics [width=0.90\columnwidth]{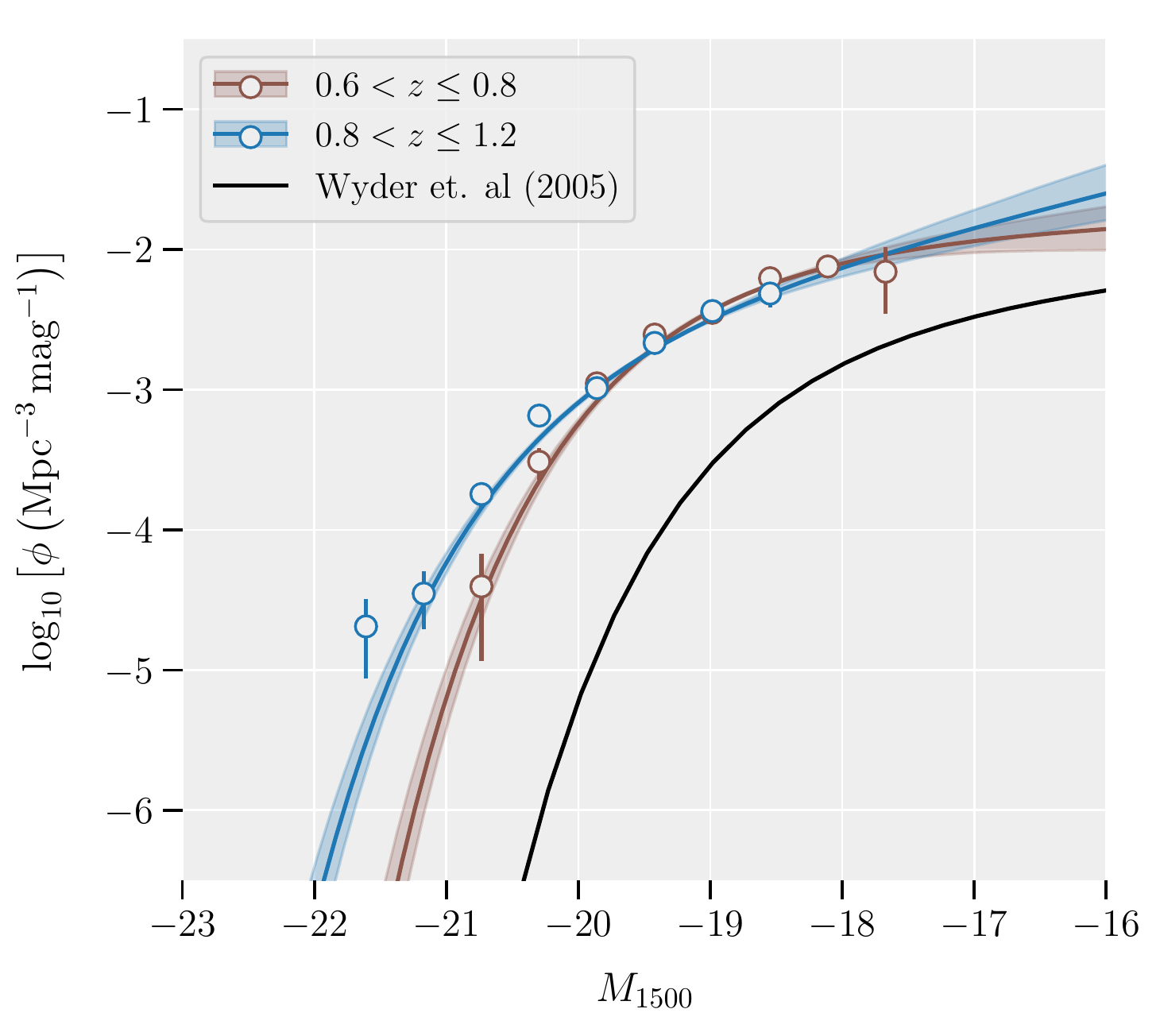}
  \caption{The estimated LF in this work. 
  The UV LF at redshift $z = 0.7$ is shown in brown colour and the 
  UV LF at redshift $z = 1$ is shown in blue.
  The solid lines, data points and the shaded areas have the same meaning as in Fig. \ref{fig:lf68}. The black solid line shows the LF estimate in the local Universe from \citet{2005ApJ...619L..15W}.}
  \label{fig:lf_comp}
\end{figure}

\section{Cosmic Variance}
\label{sec:7}
The large-scale variations in the underlying
density field of the Universe, are a major source of uncertainty in LF calculations. 
A higher or lower number count of galaxies will be observed in over dense or under dense regions of the Universe. This fluctuation will 
affect any one-point statistic based on these number counts
\citep{2004ApJ...600L.171S}.
This so called cosmic variance can significantly affect
the number density of sources for surveys exploring a small
volume of the Universe \citep{2004ApJ...600L.171S,2011ApJ...731..113M}.
Using a simple mass to light ratio, \citet{2008ApJ...676..767T} show 
that the effects of cosmic variance are not limited to number density, it can also be a source of additional uncertainty in calculations of
characteristic magnitude of a luminosity function.
We estimate the effects of cosmic variance on our calculations here.

We estimate the errors in the space density of UVW1 sources by using 
the prescription provided by \citet{2011ApJ...731..113M}, which 
calculates the cosmic variance as function of the projected sky-area,
redshift, the redshift bin size and stellar masses.
We obtain the stellar masses for our galaxies by matching our
source-list to the CANDELS-GOODS-S stellar mass catalogue from  \citet{2015ApJ...801...97S}. This survey covers 43 percent of our 
UVW1 image and contains 47 percent of our sources in the redshift
range $0.6-1.2$ using a matching radius of 1 arcsec.
We find an average stellar mass of $\SI{5.7e9}{M_\odot}$ and 
$\SI{6.0e9}{M_\odot}$ for redshift bins centered around mean redshift
0.7 and 1.0 respectively.
For these average stellar masses the average fractional cosmic variance obtained 
from the \citet{2011ApJ...731..113M} method is 0.194 for redshift bin
$0.6-0.8$ and 0.132 for redshift bin $0.8-1.2$.
We want to remark here that estimation is based on the assumption that
the mean stellar masses calculated from 47 percent of the sources
are representative of all the sources in the redshift range we are 
exploring.

There is another tool developed by \citet{2008ApJ...676..767T}, which can be used to estimate the effects of cosmic variance. 
Their work calculates cosmic variance using a different approach, 
wherein the average bias of the sample is calculated using the number density of the sources in synthetic surveys obtained from numerical N-body simulations. 
We assume $\sigma_{8}$ and an average halo occupation fraction values
of 0.8 and 0.5 respectively. In addition to these parameters we use the \citet{1999MNRAS.308..119S} bias formalism in the \citet{2008ApJ...676..767T}
web tool, and obtain $1 \sigma$ fractional errors of 0.138 and 0.104
on our normalisation for redshift 0.7 and 1.0 respectively.
The cosmic variance errors on the parameters, calculated using tools from both \citet{2008ApJ...676..767T} and \citet{2011ApJ...731..113M} are tabulated in Table \ref{tab:cv}.

In their work, \citet{2008ApJ...676..767T} also report the dependence of $M^{*}$ on the density of the large scale environment. This dependence has 
important implication when comparing the shapes of the LFs in different redshift bins or contrasting the LF shapes from different surveys.
We look at the implications of the cosmic variance on our results in the
discussion section \ref{sec:8}.

\begin{table}
  \centering
  \caption{Fractional errors on normalisation due to cosmic variance.}
  \label{tab:cv}
  \begin{tabular}{ccc}
    \hline\hline
    \noalign{\vskip 0.5mm}
    z &
    \multicolumn{2}{c}{$\Delta \phi^{*} (1 \sigma)$} \\
    \cline{2-3}
    \noalign{\vskip 0.5mm}
    &
    \citet{2011ApJ...731..113M} &
    \citet{2008ApJ...676..767T} \\
    \noalign{\vskip 0.5mm}
    \hline
    \noalign{\vskip 0.5mm}
    $0.6-0.8$              & 0.194  & 0.138  \\
    $0.8-1.0$              & 0.132  & 0.104  \\
    \hline
  \end{tabular}\\
\end{table}

\section{UV luminosity density}
\label{sec:8}

The luminosity function parameters can be strongly covariant, and 
may also depend upon the assumptions made. The integral of the UV luminosity function i.e. the total UV flux per unit comoving volume -
UV luminosity density is a much more robust quantity which can be directly translated into star formation rate density.
It can be defined as,
\begin{equation}
  \rho =  \int_{0}^{\infty} L\, \phi\left(L/L^{*}\right)\,
  \mathrm{d}\left(L/L^{*}\right),
  \label{eqn:rho}
\end{equation}
where $\phi\left(L/L^{*}\right)$ is the Schechter function 
parametrized in terms of luminosity instead of magnitudes (equation
\ref{eqn:phi}). For galaxies brighter than $L$, the luminosity density
can be calculated from the Schechter function parameters as,
\begin{equation}
  \rho = \phi^{*}\, L^{*}\, \Gamma(\alpha+2, L/L^*).
\end{equation}
We calculate the values within a range of luminosities corresponding to  $M_{1500} = [-10, -24]$.
The resulting values are tabulated in Table \ref{tab:par_f} and also plotted in Fig. \ref{fig:lum_dens} along with estimates from previous
studies.

\section{Results}
\label{sec:9}

We derived galaxy rest-frame UV LFs in the redshift range 
$0.6 \leq z < 0.8$ and
$0.8 \leq z < 1.2$, using the \cite{2000MNRAS.311..433P} method explained
in section \ref{sec:6.1}. Schechter functions were fitted to all the data to obtain 
the LF parameters as described in section \ref{sec:6.2}. The results are plotted in Fig. \ref{fig:lf68}, 
along with the LFs obtained from \citet{2005ApJ...619L..43A},  \cite{2015ApJ...808..178H} 
and \cite{2021MNRAS.506..473P} for comparison at the same redshifts. The left and 
the right 
panels in Fig. \ref{fig:lf68} show LFs derived at the mean redshift $z = 0.7$ and $z = 1.0$ respectively. 
The faintest $M_{1500}$ magnitudes explored by our UV LF estimates are 
more than one magnitude 
fainter than the characteristic magnitude $M^{*}$ in both redshift bins.

\begin{figure*}
    \centering
    \hspace*{-0.2cm}\includegraphics[width=1.2\columnwidth]{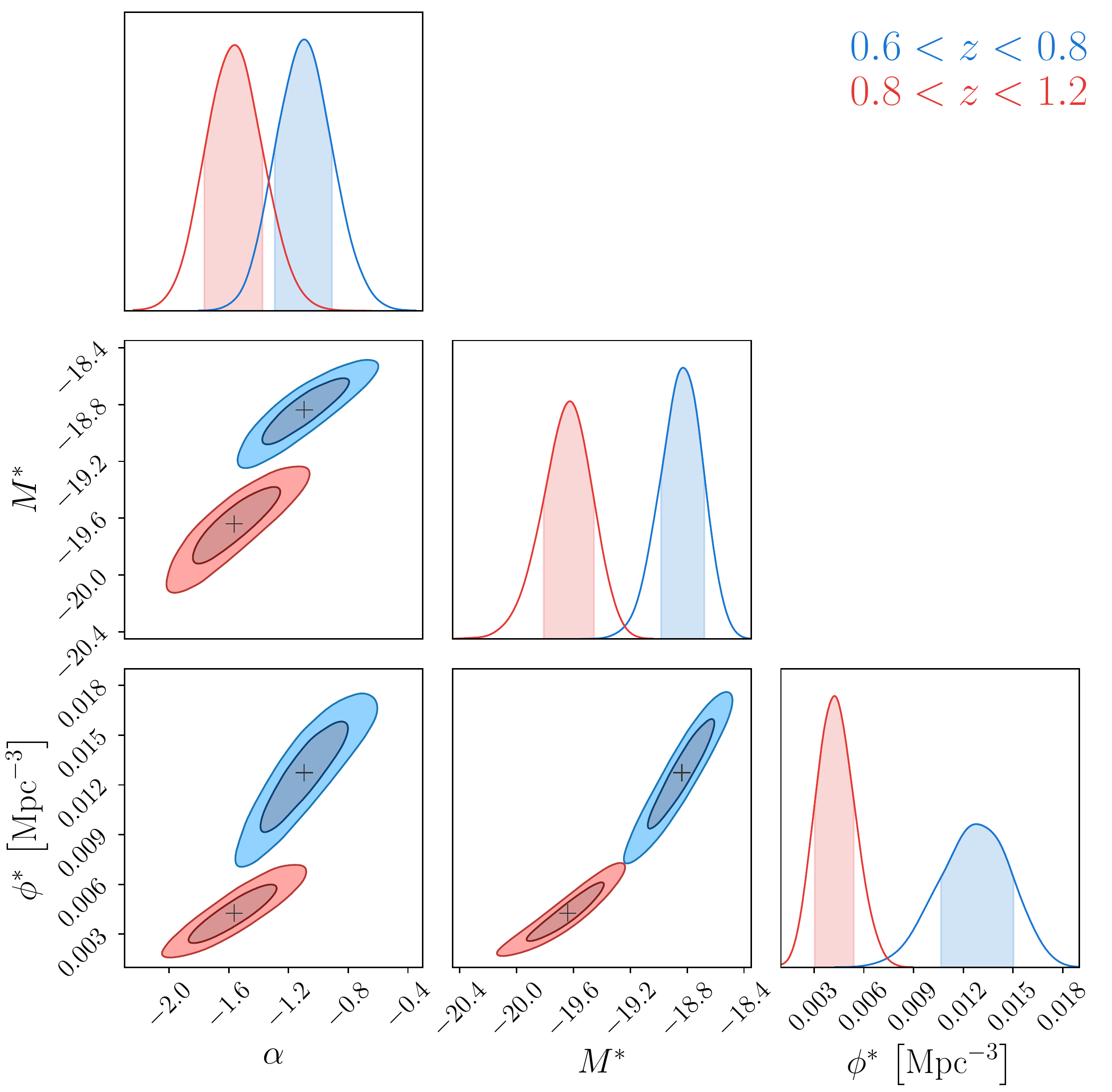}
    \caption{
    This figure represents the marginalized one dimensional (along the diagonal)
    and two dimensional (off-diagonal) posterior distributions of Schechter
    function parameters $\alpha$, $M^{*}$ and  and $\phi^{*}$. 
    The redshift bin $0.6 \leq z < 0.8$ is represented by blue and redshift bin 
    $0.8 \leq z < 1.2$ is shown in red colour. 
    The shaded region in the 
    dark and light coloured areas in the off-diagonal 
    plot correspond respectively to 68 and 95 per cent confidence 
    intervals for LF parameters.
    The black `+' symbols represent the median values for $\alpha$, $M^{*}$
    and $\phi^{*}$. The shaded region in the diagonal plots represent the one dimensional 68 per cent confidence region.}
    \label{fig:ci}
\end{figure*}

We compare the galaxy UV LF obtained at redshifts 0.7 and 1 in Fig. 
\ref{fig:lf_comp}. 
For comparison we have plotted the local luminosity function obtained by \citet{2005ApJ...619L..15W}.
Fig. \ref{fig:ci} shows the one and two dimensional 
posterior distributions
for the LF parameters, obtained by MCMC simulations.
The dark and light shaded regions show the 68 and 95 per cent confidence 
regions for three parameters.
The best fit values obtained using the maximum likelihood method presented in 
section \ref{sec:6.2} are listed in Table \ref{tab:lf_bins}.
A strong correlation can be seen in the Schechter function parameters in Fig. \ref{fig:ci}. The closed contours imply that we have sufficient data, going to faint enough magnitudes, to explore a significant 
fraction of the parameter space and constrain all the parameters.

\begin{table*}
  \setlength{\tabcolsep}{12pt}
  \centering
  \caption{Derived median Schechter function parameters of the galaxy UV LF (normalisation, characteristic magnitude and faint-end slope) from their respective posterior distributions, and the estimated value of the Luminosity density.
  The LF parameters are defined in equation \ref{eqn:phi} in section \ref{sec:4.2}, and the LD in equation \ref{eqn:rho}.
  $N_\mathrm{G}$ is the number of galaxies in each bin. Errors indicate
  $1 \sigma$ (68.26 per cent) uncertainties.}
  \label{tab:lf_bins}
  \begin{tabular}{lcccrccc}
    \hline\hline
    \noalign{\vskip 0.5mm}
    $\langle z\rangle$ &
    $z_\mathrm{min}$ &
    $z_\mathrm{max}$ &
    $N_\mathrm{G}$ &
    \multicolumn{1}{c}{$\phi^* / 10^{-3}$} &
    \multicolumn{1}{c}{$M^*$} &
    \multicolumn{1}{c}{$\alpha$} &
    \multicolumn{1}{c}{$\rho / 10^{26} $} \\
    &
    &
    &
    &
    \multicolumn{1}{c}{(Mpc$^{-3}$)}
    & 
    &
    &
    \multicolumn{1}{c}{(erg s$^{-1}$ Hz$^{-1}$ Mpc$^{-3}$)}\\
    \hline
    \noalign{\vskip 0.5mm}
    0.7$^a$ & 0.60 & 0.80 & 545 & $12.73^{+2.03}_{-2.25}$ & $-18.84^{+0.14}_{-0.15}$ & $-1.10^{+0.19}_{-0.18}$ &
    $2.02^{+0.26}_{-0.18}$ \\
    \noalign{\vskip 0.5mm}
    1.0$^a$ & 0.80 & 1.20 & 534 & $4.26^{+1.18}_{-1.12}$ & $-19.64^{+0.16}_{-0.18}$ & $-1.56^{+0.19}_{-0.18}$  &
    $2.63^{+1.04}_{-0.55}$ \\
    \noalign{\vskip 0.5mm}
    \hline
  \end{tabular}\\
  \begin{minipage}{0.92\textwidth}
      \textsuperscript{$a$}{Average redshift for the bin.}
  \end{minipage}
  \label{tab:par_f}
\end{table*}

It can be seen from Table \ref{tab:lf_bins} that our best fit
value of the faint-end slope changes by $1.85 \sigma$ as redshift increases from $z = 0.7$ to $z = 1$, 
while the characteristic magnitude evolves significantly.
A brightening of $0.8$ mag at $> 3 \sigma$
in $M^{*}$ is observed as the redshift increases.
This is also evident in Fig. \ref{fig:lf_comp} 
where the $1 \sigma$ error 
regions corresponding to both LF just overlaps towards the fainter magnitudes implying an insignificant change, whereas at
the bright end the error regions are well separated, and the 
LF (blue coloured) representing the redshift bin centered at $z = 1$
extends towards brighter magnitudes as compared to the 
other (brown) LF for $z = 0.7$.
The normalisation $\phi^{*}$, changes to become
three times its value at $z = 1$ as we move to $z = 0.7$ 
(Fig. \ref{fig:ci}).
It is very important to remark here that the effects of cosmic variance
have to be kept in mind while looking at the changes in the LF parameter $\phi^*$ from one redshift bin to another.

The bright ends of the LFs ($M_{1500} < 21$ mag for $z = 0.7$ and 
$M_{1500} < 22$ mag for $z = 1$) are not very well constrained.
We note here that to constrain this regime of the LF properly, we need 
observations from a very large area of the sky
to have more bright sources in the sample.

\begin{figure}
    \centering
    \hspace*{-0.3cm}\includegraphics[width=\columnwidth]{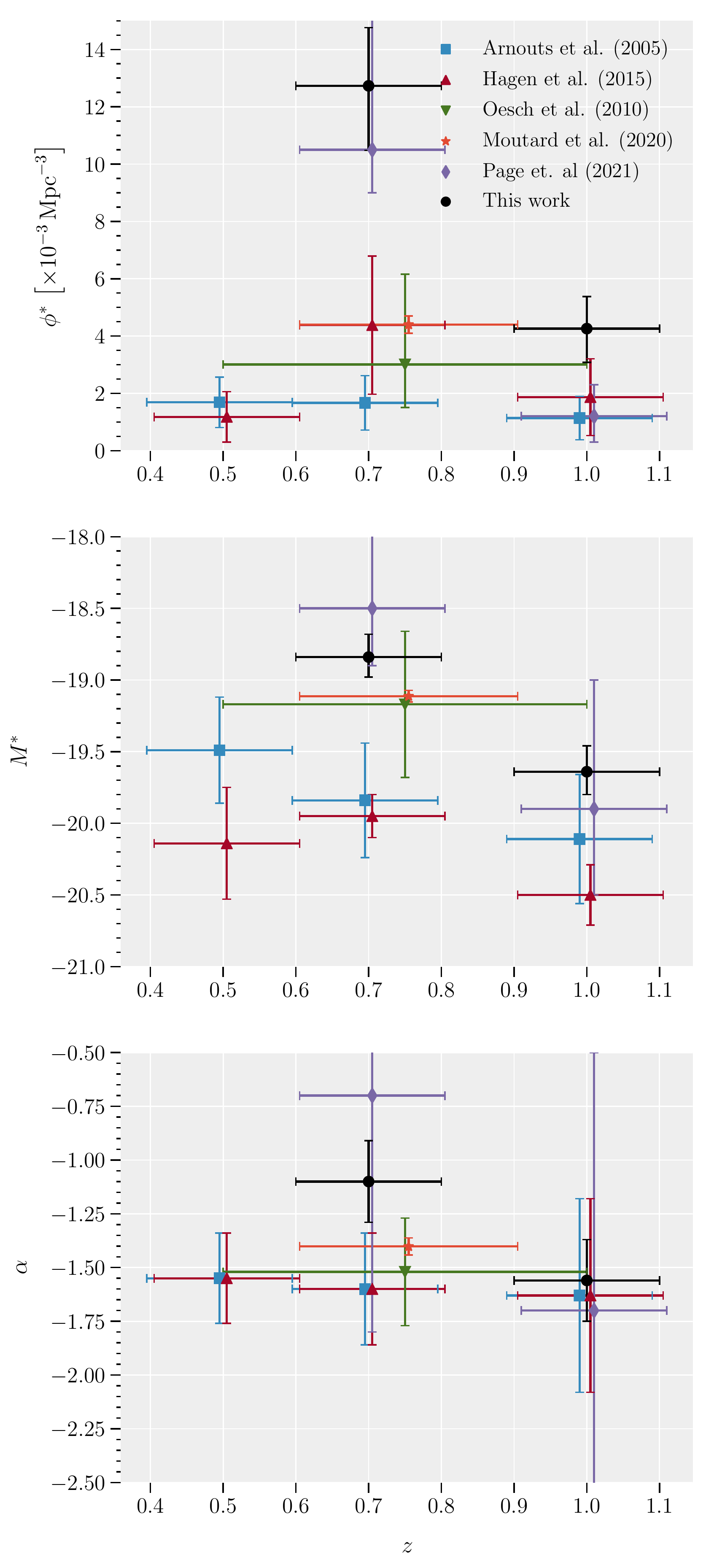}
    \caption{The Schechter function parameters $\alpha$, $M^{*}$ and 
    $\phi^{*}$.
    The values estimated from this work are in black colour and other
    values from \citet{2005ApJ...619L..43A,2015ApJ...808..178H,2010ApJ...725L.150O,2020MNRAS.494.1894M} and 
    \citet{2021MNRAS.506..473P} are in blue, red, green, orange and purple  colours 
    respectively.
    These parameters are defined in Equation \ref{eqn:phi}.
    The top panel represents the characteristic number density $\phi^{*}$, 
    central panel shows characteristic magnitude $M^{*}$
    and the bottom panel shows the 
    variation in faint-end slope $\alpha$ with redshift.
    The horizontal error bars represent the width of the redshift bin and 
    the vertical ones represent one $\sigma$ (68.26 per cent) uncertainties. 
    These values are also
    tabulated in Table \ref{tab:comparison_t}.
    The redshifts on the horizontal axis are perturbed very slightly to clearly show the the data points separate from each other.}
    \label{fig:comp_f}
\end{figure}

\begin{table*}
  \setlength{\tabcolsep}{15pt}
  \caption{The derived values of the Schechter function parameters
  {$\phi^*$}, {$M^*$} and {$\alpha$} from previous studies.}
  \label{tab:comparison_t}
  \begin{tabular}{lllrr}
    \hline\hline
    \noalign{\vskip 0.5mm}
    \multicolumn{1}{l}{$\text{Work}$} &
    $\langle z\rangle^{a}$ &
    \multicolumn{1}{c}{$\phi^*$/$10^{-3}$} &
    \multicolumn{1}{c}{$M^*$} &
    \multicolumn{1}{c}{$\alpha$} \\
    &
    &
    \multicolumn{1}{c}{(Mpc$^{-3}$)}
    & \\
    \hline
    \noalign{\vskip 0.5mm}
    \citet{2007ApJ...654..172D} & 0.55 
    & $6.23^{+2.68}_{-2.12}$
    & $-19.22^{+0.28}_{-0.28}$ & $-1.39^{+0.14}_{-0.13}$ \\ 
    \noalign{\vskip 0.5mm}
    \citet{2005ApJ...619L..43A} & 0.7 
    & $1.67^{+0.95}_{-0.95}$ 
    & $-19.84^{+0.40}_{-0.40}$ & $-1.60^{+0.26}_{-0.26}$ \\
    \noalign{\vskip 0.5mm}
    \citet{refId0} & 0.7 
    & $9.53^{+0.99}_{-0.99}$ 
    & $-18.30^{+0.10}_{-0.10}$ & $-0.90^{+0.08}_{-0.08}$\\
    \noalign{\vskip 0.5mm}
    \citet{2015ApJ...808..178H} & 0.7 
    & $6.65^{+1.21}_{-1.21}$ 
    & $-19.78^{+0.10}_{-0.10}$ & $-1.60^{b}$\\
    \citet{2021MNRAS.506..473P} & 0.7 
    & $10.5^{+1.20}_{-5.20}$ 
    & $-18.50^{+0.40}_{-0.60}$ & $-0.70^{+1.10}_{-1.10}$\\
    \noalign{\vskip 0.5mm}
    \citet{2010ApJ...725L.150O} & 0.75 
    & $3.01^{+1.50}_{-3.15}$ 
    & $-19.17^{+0.51}_{-0.51}$ & $-1.52^{+0.25}_{-0.25}$ \\
    \noalign{\vskip 0.5mm}
    \citet{2014ApJ...794L...3W} & 0.75 
    &  $3.56^{+5.25}_{-1.79}$ 
    & $-18.62^{+0.27}_{-0.21}$ & $-1.23^{+0.14}_{-0.07}$ \\
    \citet{2020MNRAS.494.1894M} & 0.75 
    &  $4.40^{+0.30}_{-0.30}$ 
    & $-19.11^{+0.04}_{-0.04}$ & $-1.40^{+0.04}_{-0.04}$ \\
    \noalign{\vskip 0.5mm}
    \citet{refId0} & 0.9 
    &  $9.01^{+0.94}_{-0.96}$ 
    & $-18.70^{+0.10}_{-0.10}$ & $-0.85^{+0.10}_{-0.10}$ \\
    \noalign{\vskip 0.5mm}
    \citet{2005ApJ...619L..43A} & 1.0 
    & $1.14^{+0.76}_{-0.76}$ 
    & $-20.11^{+0.45}_{-0.45}$ & $-1.63^{+0.45}_{-0.45}$ \\
    \noalign{\vskip 0.5mm}
    \citet{2015ApJ...808..178H} & 1.0 
    & $1.36^{+0.10}_{-0.10}$ 
    & $-20.74^{+0.12}_{-0.12}$ & $-1.63^{b}$ \\
    \citet{2021MNRAS.506..473P} & 1.0 
    & $1.20^{+0.90}_{-1.10}$ 
    & $-19.90^{+0.60}_{-0.90}$ & $-1.70^{+1.20}_{-0.80}$\\
    \noalign{\vskip 0.5mm}
    \citet{refId0} & 1.1 
    & $7.43^{+1.08}_{-1.15}$ 
    & $-19.00^{+0.20}_{-0.20}$ & $-0.91^{+0.16}_{-0.16}$ \\
    \noalign{\vskip 0.5mm}
    \hline
  \end{tabular}\\
  \begin{minipage}{0.73\textwidth}
    \textsuperscript{$a$}{Average redshift for the bin.}\\
    \textsuperscript{$b$}{These faint-end slope values are fixed to those obtained by \citet{2005ApJ...619L..43A}.}\\
  \end{minipage}
\end{table*}

\section{Discussion}
\label{sec:10}

We compare the outcomes of our analysis with the literature values of
the Schechter function parameters in
the redshift range $0.6 \leq z < 0.8$ and $0.8 \leq z < 1.2$. 
For comparison we tabulate 
previous estimates in Table \ref{tab:comparison_t} and plot some of
these values in Fig. \ref{fig:comp_f}. 
For mean redshift $z = 0.7$ we find
that our results lie just at the edge of the $2 \sigma$ region around values
reported by \citet{2005ApJ...619L..43A}.
The faint-end slope reported by their study is $\alpha = -1.60 \pm 0.26$, 
whereas we get a flatter value $\alpha = -1.10^{+0.19}_{-0.18}$. As can
be seen  
in the left panel of Fig. \ref{fig:lf68}, there is a deviation from 
these works at
the bright-end as well, however we note here that a survey wider than 
the 
CDFS with more bright galaxies is required to constrain the very bright end of the LF. 
The best-fit characteristic magnitude $M^{*}$ resulting from our 
analysis is about a magnitude fainter than what is inferred by \citet{2005ApJ...619L..43A}.
Thanks to the exceptionally long total exposure time of our XMM-OM UVW1 image,
we manage to get the best constraints to date on the faint-end slope, from 
a survey observing directly in the rest frame 1500 {\AA} at these redshifts.
Compared to \citet{2005ApJ...619L..43A}, this study improves the 
$1 \sigma$
constraints on $\alpha$ by 44 per cent on averaging both redshift bins, whereas the $M^{*}$ 
uncertainties are reduced by a significant margin of 63 per cent on average.

\begin{figure}
  \hspace*{-0.3cm}\centering
  \includegraphics [width=0.95\columnwidth]{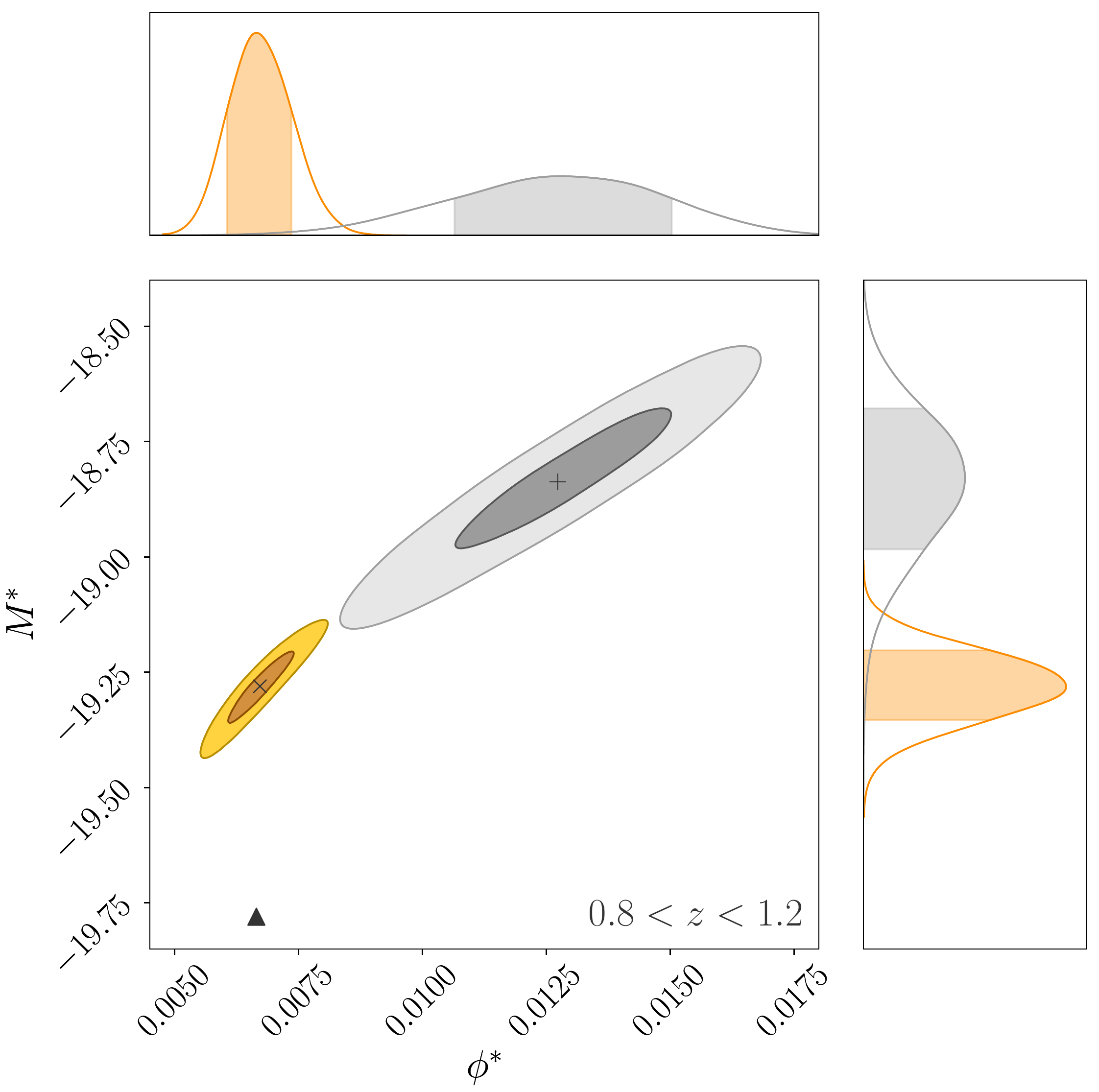}
  \includegraphics [width=0.95\columnwidth]{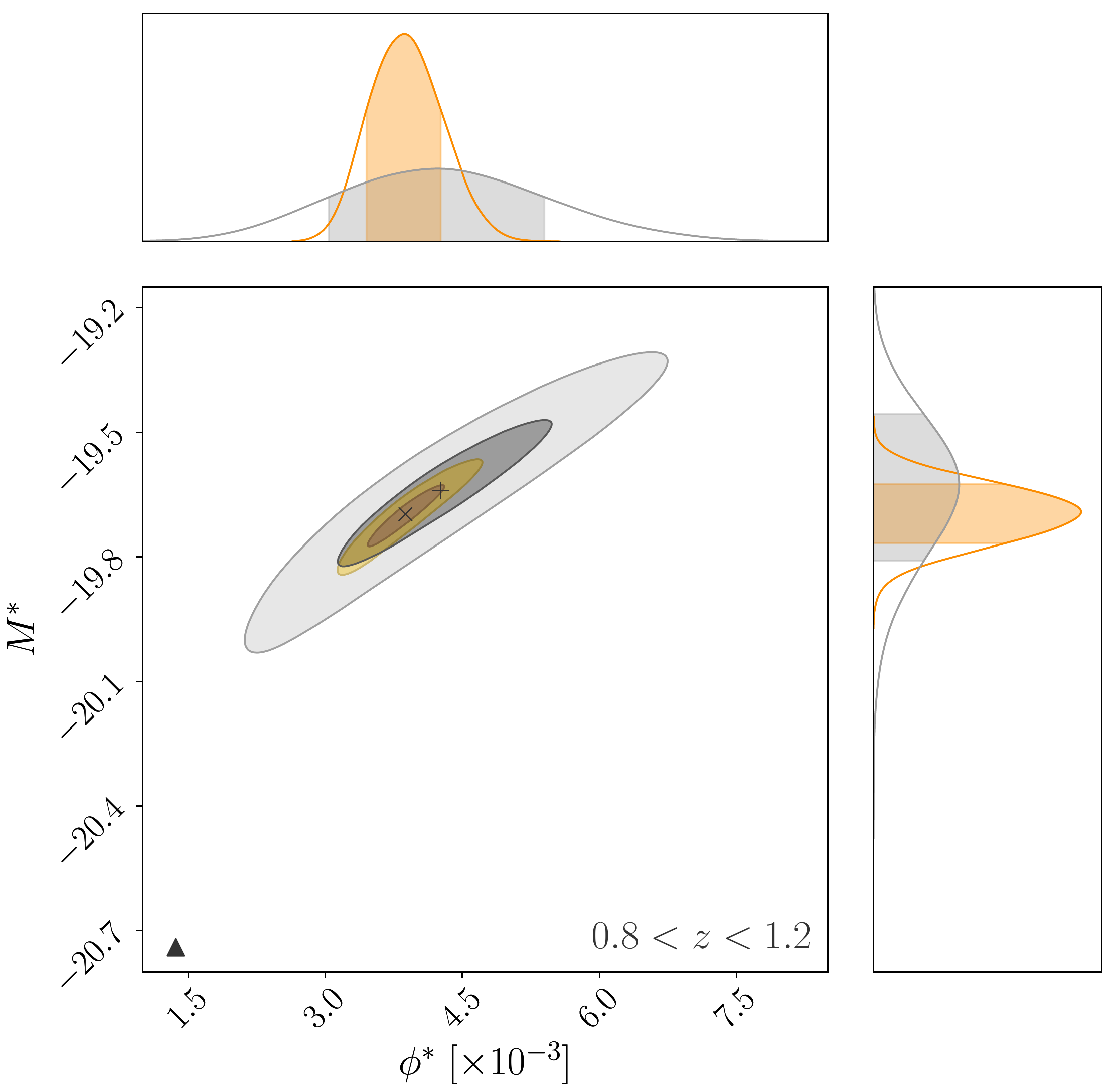}
  \caption{Here we plot the estimates for $M^{*}$ and $\phi^{*}$.
  The top and bottom panels represent the redshift bins $0.6 < z < 0.8$
  and $0.8 < z < 1.2$ respectively. `+' symbols within the gray contours represent the values obtained
  from this work when we treat $\alpha$ as a free parameter. `$\times$'
  symbols in the yellow contours represent the values from this work if 
  we fix the faint end slope to \citet{2005ApJ...619L..43A}. The values 
  obtained by \citet{2015ApJ...808..178H} are shown as the upward triangle. The definitions of the contours are the same as in Fig. \ref{fig:ci}.}
  \label{fig:corn_hag}
\end{figure}

We compare our results to \citet{2015ApJ...808..178H}, who used the observations
from the same field as ours (i.e. CDFS) for their calculations, and report smaller uncertainties on $M^{*}$. It should be noted that they used U
band observations from \textit{Swift}-UVOT to select their sources. This affects the minimum 
limiting magnitude in the rest-frame FUV band, and so the data used 
for their UV LF
calculations do not go deep enough in the rest-frame FUV to put constraints on
the faint-end slope. They fixed their faint-end slope equal to the 
\citet{2005ApJ...619L..43A} values to calculate other parameters. A correlation between 
faint-end slope and characteristic magnitude, demands their $M^{*}$ estimates
to be strongly dependent on \citet{2005ApJ...619L..43A}. We want to remark here that our data reach $\simeq$ 0.5 mag and $>$ 1.0 mag deeper than \citet{2005ApJ...619L..43A} and \citet{2015ApJ...808..178H} respectively in this redshift bin.
In Fig. \ref{fig:corn_hag}, we plot our parameter estimates for $M^{*}$
and $\phi^{*}$ after fixing the faint end slope values to the ones
from \citet{2005ApJ...619L..43A}, similar to \citet{2015ApJ...808..178H}.
We see the confidence contours for the fixed $\alpha$ (in yellow) shrink dramatically as compared to the ones obtained for variable
$\alpha$ (in grey) for both redshift bins. 
For redshifts $0.6 \leq z < 0.8$ (upper panel in Fig. \ref{fig:corn_hag}),
the normalisation obtained by \citet{2015ApJ...808..178H} is within
1 $\sigma$ if we fix the faint end slope to \citet{2005ApJ...619L..43A},
however there is different characteristic magnitude. This difference 
could be because of the different selection technique applied by 
\citet{2015ApJ...808..178H}.

\citet{2010ApJ...725L.150O} estimated a faint-end slope 
$\alpha = -1.52 \pm 0.25$, 
in the redshift range $0.5 < z < 1$ having a mean redshift very close 
to ours. 
Their result for $\alpha$ and the break magnitude $M^{*}$ agrees with 
our values within $1 \sigma$.
The UV LF estimated by \cite{2021MNRAS.506..473P} in this redshift bin seems to 
follow the same shape as ours (Fig. \ref{fig:lf68}), with a different 
normalisation. 
They also obtain a 
faint value for the break magnitude ($M^{*} = -18.50^{+0.40}_{-0.60}$). However, due to a small sample size, their value is rather uncertain.

The faint-end slope deduced by \citet{refId0} and 
\citet{2020MNRAS.494.1894M}
in the similar redshift bin are $-0.90 \pm 0.08$ and $-1.40 \pm 0.04$
respectively. 
Thus our values in this redshift bin agree well with their values and within $1 \sigma$ and $2 \sigma$.
However, when comparing our results to these works, 
it must be pointed out here again that they obtained the UV LF by 
extrapolating their measurement to 1500 {\AA} from longer wavelengths. 

There is no tension between our faint-end slope and 
\citet{2014ApJ...794L...3W} (i.e. $-1.23 \pm 0.14$) determined for a mean redshift of $z = 0.75$, up to $1 \sigma$. 
This study used a very different method, galactic archaeology of the 
galaxies in the local group, to estimate the UV LF. They were able to 
estimate the star formation history and hence the LFs of galaxies to 
$z \sim 5$, probing very faint magnitude limits ($M_{V} = -4.9$)

In the redshift range $0.8 \leq z < 1.2$, again our data is $> 1.0$ 
mag 
deeper compared with \citet{2015ApJ...808..178H}. With respect to \citet{2005ApJ...619L..43A} our data is $\sim 0.15$ mag deeper.
Overall our values of the UV LF parameters in this redshift bin are in accordance
with findings reported by \citet{2005ApJ...619L..43A}, with better constraints on the faint-end slope.
Their estimates of $\alpha = -1.63 \pm 0.45$ 
are within $1 \sigma$ from our estimate $\alpha = -1.56^{+0.19}_{-0.18}$.
The value of the characteristic magnitude obtained in our study,
$M^* = -19.64^{+0.16}_{-0.18}$, 
is in reasonable agreement with the one determined by
\citet{2005ApJ...619L..43A}, $M^* = -20.11 \pm 0.45$, considering the
errors. 
The bright-end of the LF in this redshift bin does not match
with \citet{2015ApJ...808..178H}, and their $M^{*}$ values are beyond
$1 \sigma$ of our estimate. 
The situation does not change much if we 
fix the faint-end slope to their values.
From the lower panel of Fig. \ref{fig:corn_hag}, one can see 
that the results obtained from our data for a fixed and variable
faint-end slope match very well. This is because 
our fixed faint-end slope was very close to the median
of our marginalised posterior distribution of $\alpha$. 
But, despite a fixed value of $\alpha$,
the normalisation and characteristic magnitude do not match with
\citet{2015ApJ...808..178H}, and again, we think the reason could be
the different methods applied to select the sources.
There is considerable discrepancy between our best-fit value of $\alpha$ and $M^{*}$ and those obtained by \citet{refId0} 
at $z = 0.9$ and $z = 1.1$.

As apparent from Table \ref{tab:comparison_t} the best-fitting faint-end slope values in the literature
can be as steep as $-1.70$ and as shallow as $-0.85$. This variation in the values of $\alpha$ represents a very substantial level of uncertainty in the characterisation of galaxy UV LFs. 
We obtain a shallow faint-end slope for redshift $z = 0.7$,
compared to estimates from \citet{2010ApJ...725L.150O} and
\citet{2005ApJ...619L..43A}.
The faint-end slope
reflects the contribution of the low luminosity galaxies to the LF, 
so, having a smaller value than previous studies suggests that there were
fewer faint galaxies at this epoch in the history of the Universe than estimated by previous
studies. Our data reach $1.2$ mag fainter than the characteristic
magnitude $M^{*}$, so, we are confident that the faint end is sampled
well.
Many previous works conclude a constant value of the faint-end slope
between average redshift 0.7 to 1. Our values of $\alpha$ agree with
this picture as we see very modest
(close to $2 \sigma$) change in $\alpha$ as redshift increases to 
$z = 1$.

The uncertainties in $\alpha$ are also followed by uncertainties in 
characteristic magnitude $M^{*}$ (see Fig. \ref{fig:comp_f} and Table \ref{tab:comparison_t}), because of the correlation between these two 
parameters as apparent from Fig. \ref{fig:ci}. 
The characteristic magnitude $M^{*}$ estimated in this work is fainter 
than those reported by \citet{2015ApJ...808..178H} and 
\citet{2005ApJ...619L..43A}, at both redshift bins.
We see an evolution in characteristic magnitude with 
redshift in the range we explored, which can be attributed to evolution 
of the population towards brighter objects as we get closer to the peak 
of the SFR in the Universe. 
As mentioned earlier cosmic variance needs to considered while making any
claims of the evolution of the LF parameters with redshift. 
Because of the presence of over-dense
regions in the CDFS \citep{2003ApJ...592..721G,2014AJ....147...52D}, which lie in our $0.6-0.8$ redshift bin, the space density and hence the number 
count of the galaxies is higher in this bin than the average of other surveys.
As shown by \citet{2008ApJ...676..767T}, regions with above-average space 
density are biased towards brighter values of $M^{*}$. 
So, the underlying value of characteristic magnitude should be fainter than 
our estimated value in this redshift range. This should only make the evolution in the $M^{*}$
between the bins more significant.
It is also important to mention that \citet{2008ApJ...676..767T} have not
considered baryonic feedback effects in their modelling, and their 
results may change if these effects are included.
A large change in normalisation $\phi^{*}$ value is seen with change in
redshift, which can again be attributed to the large galaxy clusters 
in the redshift bin centered at $z = 0.7$. We would like to direct 
the reader's attention towards Table \ref{tab:cv}, which shows the relative error in the value
of $\phi^{*}$ calculated using methods suggested by \citet{2008ApJ...676..767T} and \citet{2010ApJ...710..903M}. These errors
should be considered in addition to the total statistical errors
quoted with the parameter values in Table \ref{tab:par_f}.

We compare our UV LFs to the local UV LF calculated by \citet{2005ApJ...619L..15W} using \textit{GALEX} FUV and NUV
data. They obtained $-18.04 \pm 0.11$ and $-1.22 \pm 0.07$ for $M^*$ and $\alpha$ respectively for the FUV band. 
With respect to their findings our results at $z = 0.7$ suggest an evolution in $M^{*}$ best-fit values by 0.8 mag which is significant at
$4 \sigma$, from $z = 0.7$ to the present time.
This evolution can be seen in Fig. \ref{fig:comp_f}, where we plot the 
\citet{2005ApJ...619L..15W} results along with our results.
As far as $\alpha$ is concerned we do not see any significant change
between $z = 0.7$ and the local value.

\begin{figure}
  \hspace*{-0.3cm}\centering
  \includegraphics [width=\columnwidth]{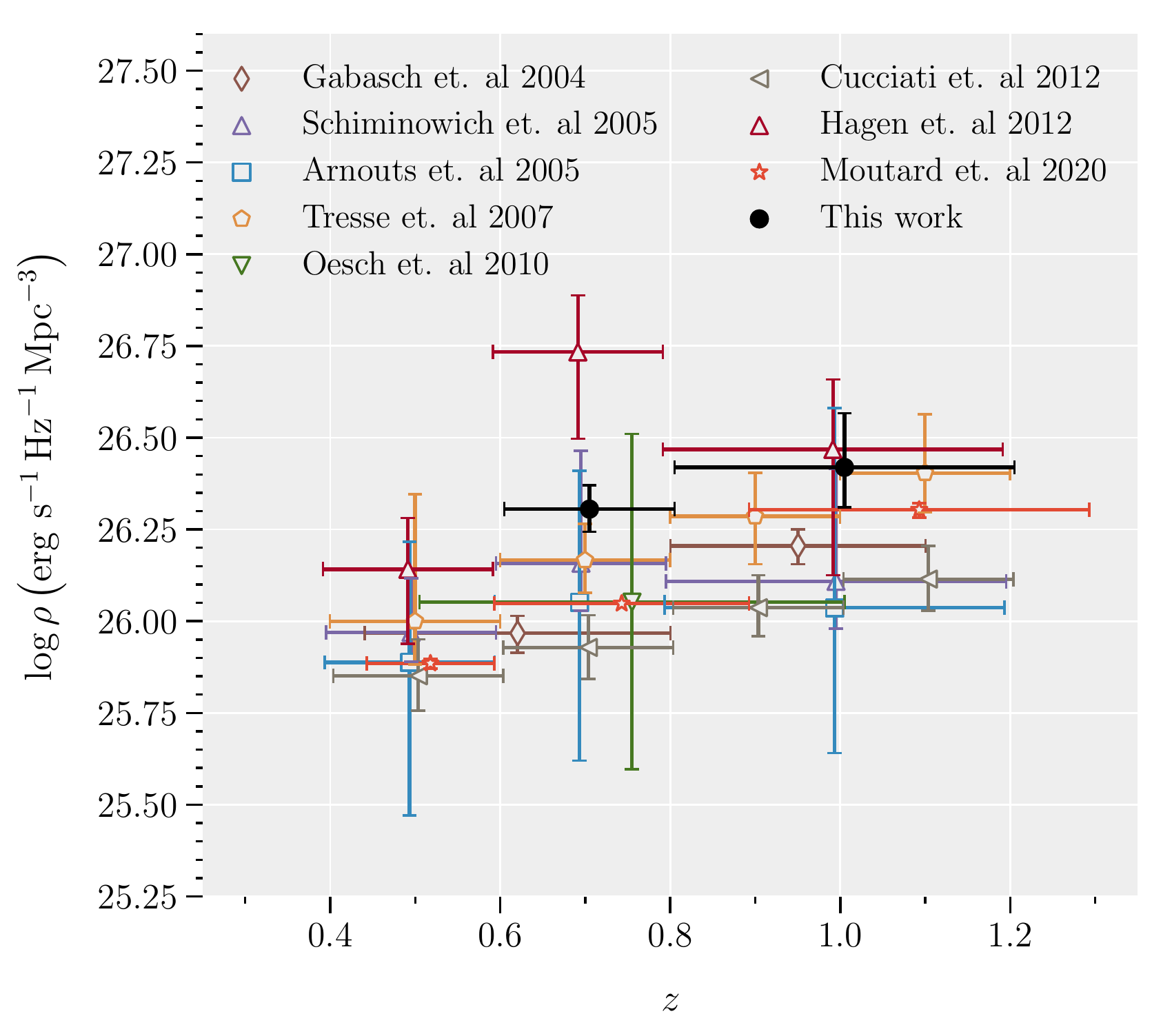}
  \caption{The luminosity density estimated by integrating the luminosity function. The data points in different colours represent the observed luminosity calculations from the following studies : \citet{2004A&A...421...41G,2005ApJ...619L..47S,2005ApJ...619L..43A,2007A&A...472..403T,2010ApJ...725L.150O,refId0,2015ApJ...808..178H,2020MNRAS.494.1894M}. Our estimates are shown in black colour. The vertical and horizontal error bars represent the 1 $\sigma$ (68.26 per cent) uncertainties and the redshift bin edges respectively. For clarity the data points at same redshifts are slightly moved.}
  \label{fig:lum_dens}
\end{figure}

The luminosity density of the Star forming galaxies is calculated by integrating the UV LF, and plotted in Fig. \ref{fig:lum_dens} along with values from other works. Our results for the luminosity density fall within the error regions of past studies,
and follow the same trend as the redshift changes i.e. higher luminosity density at redshift $z=1$ as compared to its value at
redshift $z=0.7$.

We consider the effect of potential AGN contribution to our UV LF calculations. In order to do that, 
we create two source-lists. One by removing all
potential AGN (these sources include all the sources identified by
\citet{2016ApJS..224...15X} or \citet{2017ApJS..228....2L} as AGN and all sources
with $0.5 - 8$ KeV X-ray luminosity higher than $10^{42}$ ergs s$^{-1}$)
from the sample. 
The second by including all these (potential AGN) sources in the final source-list.
The distributions of the resulting two source-lists as a function of their $M_{1500}$
magnitudes are plotted in Appendix~\ref{sec:agn} to show the 
contribution of these sources at the bright end of the luminosity function.
We compare the Schecter function parameters obtained from both cases.
The best-fit
results for the case 
where all the identified AGN and bright X-ray sources are removed, stay within the $1 \sigma$ uncertainties of our actual result
(obtained using the X-ray flux cut only). 
However, the values of LF parameters change drastically when all 
the potential AGN are included in the sample. The exact numbers for 
both cases are tabulated in the Table \ref{tab:agn} and plotted in Fig. \ref{fig:agn} in Appendix~\ref{sec:agn}. This result could also serve as
an explanation to why we find a smaller number of bright galaxies 
as compared to \citet{2015ApJ...808..178H}. 
Even a small number of bright AGN present in their sample could brighten their characteristic magnitude by a significant amount,
and consequently increase the luminosity density (Fig. \ref{fig:lum_dens}). 

\section{Summary and Conclusion}
\label{sec:11}

We have measured the UV LF of galaxies over redshifts ranging from
$z = 0.6$ to $z = 1.2$ using data from the XMM-OM survey of the CDFS.
Our estimates of the UV LF extend to deeper magnitudes than  
previous studies, with $M_{1500} = [-20.75, -17.70]$ and
$[-21.60, -18.55]$ at $0.6 \leq z < 0.8$ and $0.8 \leq z < 1.2$, respectively.
We use Monte-Carlo simulations to estimate completeness of the data and the magnitude error 
distribution used to account for any photometric errors before the 
fitting process is executed.
Using the \citet{2000MNRAS.311..433P} method, we construct binned
realisations of the UV galaxy LF as a function of 1500  {\AA} rest 
frame magnitudes in two redshift bins with average redshifts 
$z = 0.7$ and $z = 1.0$. 

The LF is well described by the Schechter function shape. 
We calculate the best-fit values of the galaxy LF parameters by fitting
the Schechter function to the data by using parametric maximum-likelihood in 
both redshift bins. From our fits we obtain a relatively flat
value of the faint-end slope $\alpha = -1.10^{+0.19}_{-0.18}$ in the 
redshift range $0.6 \leq z < 0.8$ compared with most previous studies, 
and in the redshift bin, 
$0.8 \leq z < 1.2$ we obtain a best-fit value $\alpha = -1.56^{+0.19}_{-0.18}$.
There is a small change in $\alpha$ with low statistical significance between the two redshift bins.

Regarding the characteristic magnitude
$M_{*}$, our derived values are 0.5 to 1 mag fainter than some previous studies in the same redshift ranges. 
Between the redshift bins under consideration, $M_{*}$ also evolves by 
$0.8$ mag at $3 \sigma$. 

The value of the characteristic number density $\phi^{*}$ agrees
with one of the previous works in the redshift bin centered at $z = 0.7$.
At $z = 1$, our estimate for $\phi^{*}$ is within $2 \sigma$ of the 
closest 
value from literature. We attribute the differences to the cosmic 
variance of the different fields chosen by different works. Its values would be better constrained with estimates from different parts of the 
sky to counter the cosmic variance.
The luminosity density for our sample is in agreement with the 
previous studies.

The AGN population in a galaxy sample can bias the UV LF 
parameters, if not handled properly. An effective way to remove the AGNs
is using a total X-ray luminosity cut at $10^{42}$ ergs s$^{-1}$. 

The remarkable potential of the XMM-OM for this type of study is clear 
from the results so far. This OM data set with its observations of the 
CDFS has demonstrated already that it is a powerful tool for 
constraining the faint end of UV LF and thus exploring galaxy evolution to redshift 1. Further wide field data will provide critical insights into the bright end of the LF.

\section{Acknowledgements}
\label{sec:12}
This research makes use of the observations taken with XMM-Newton telescope, 
an ESA science mission with instruments and contributions directly funded by ESA Member States and NASA.
MJP acknowledges support from the UK Science and Technology Facility Council (STFC) grant number ST/S000216/1.
MS would like to thank Vladimir Yershov for outstanding support with 
the XMM-OM software. MS also thanks Ignacio Ferreras, Paul Kuin and 
Sam Oates for valuable comments and discussions related to this
manuscript. MS would like to extend their gratitude towards Michele Trenti
for sharing the source code for their CV calculator.
We thank the anonymous referee for their constructive report to further improve this manuscript.

\section{Data Availability}
\label{sec:13}
The data used in this article can be obtained from the \textit{XMM-Newton} Science archive (XSA) at \url{https://www.cosmos.esa.int/web/xmm-newton/xsa}. We make the source-list used in this paper as a supplementary table with the online version of the paper.
Other supporting material related to this article is available on a 
reasonable request to the corresponding author.

\bibliographystyle{mnras}
\bibliography{References}

\appendix
\section{AGN in the Sample}
\label{sec:agn}
\begin{figure}
    \centering
    \hspace*{-0.3cm}\includegraphics[width=0.95\columnwidth]{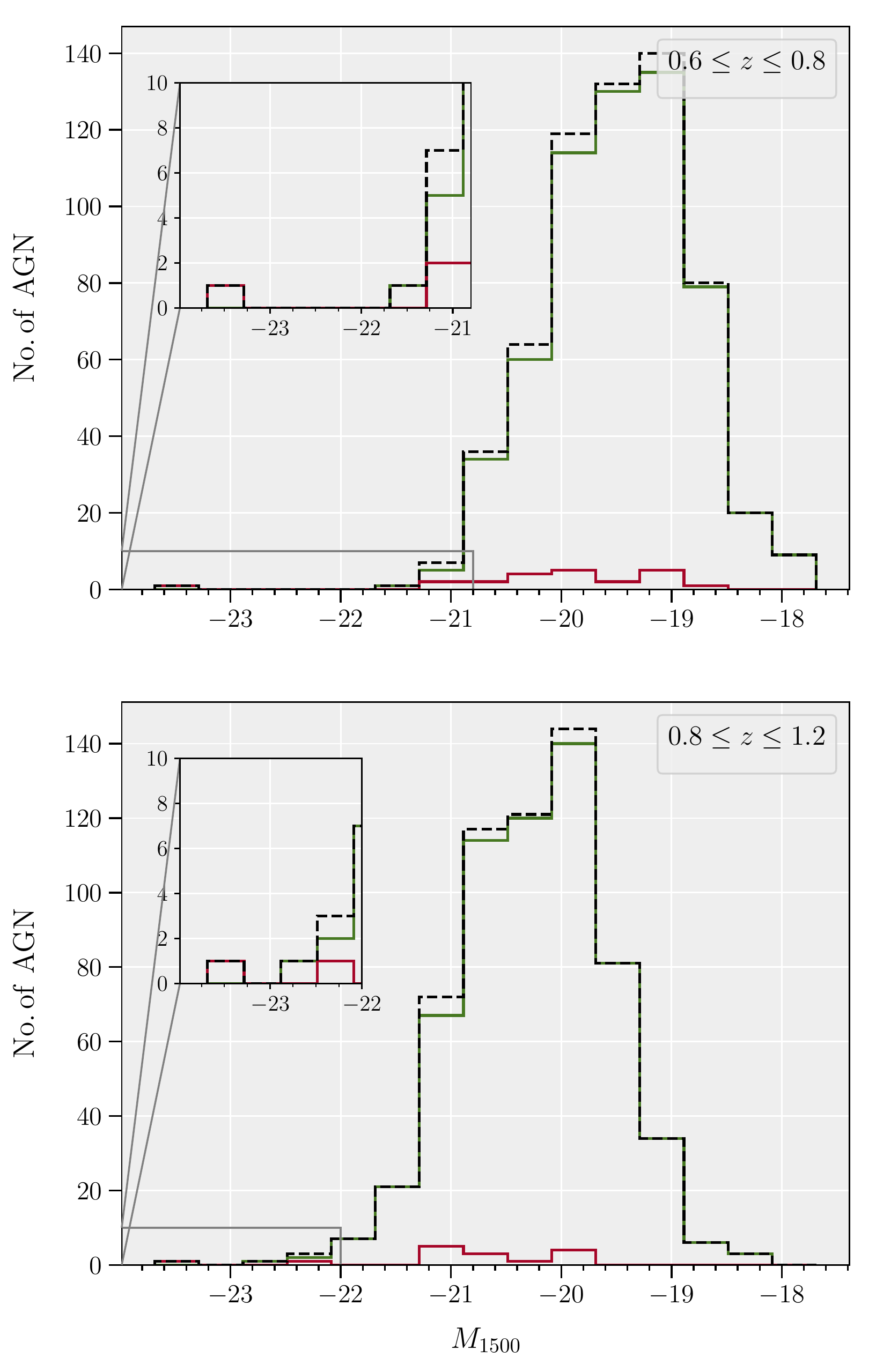}
    \caption{We show here the distribution of sources in our sample as a function of $M_{1500}$ absolute magnitude. 
    The top and bottom panels show the distribution in redshift bins 
    $z = 0.7$ and $z = 1.0$ respectively.
    The red histogram represents the X-ray sources with luminosity
    $10^{42}$ ergs s$^{-1}$ and sources identified as AGN in \citet{2016ApJS..224...15X} and \citet{2017ApJS..228....2L} X-ray catalogues from 
    CDFS. 
    The histogram in green shows the source-list with all the 
    sources in the red distribution removed.
    The distribution in black represents the total source-list including
    all the Star-forming galaxies and identified X-ray sources.
    The inset figures in each panel show the distribution at bright ends
    stretched along the y-axis.}
    \label{fig:agn}
\end{figure}

As mentioned in section \ref{sec:4.2}, the X-ray sources are identified after cross-correlating our source-list with \citet{2016ApJS..224...15X} and \cite{2017ApJS..228....2L} catalogues.
In our overall source-list of 2581 sources, 99 are identified as X-ray sources, 63 of which have a $0.5 - 8$ KeV X-ray luminosity higher than $10^{42}$ ergs s$^{-1}$. 
We only consider these 63 bright sources as the ones that may host
powerful enough AGN, such that the UV radiation coming from an accretion disc 
around their super-massive black holes dominates over any coming from the star formation activity.
These can contaminate our source-list and hamper the UV LF calculations.
17 of these bright X-ray sources fall in the $0.6 \leq z < 0.8$ redshift bin, and 
have $M_{1500}$ absolute magnitude in the range $[-23.65, -18.62]$.
In the redshift bin $0.8 \leq z < 1.2$, 11 bright X-ray sources are identified with this criteria with $M_{1500} = [-24.17, -19.95]$.
The distributions of the identified X-ray sources are plotted in Fig.
\ref{fig:agn}.

\begin{figure}
    \centering
    \hspace*{-0.3cm}\includegraphics[width=0.95\columnwidth]{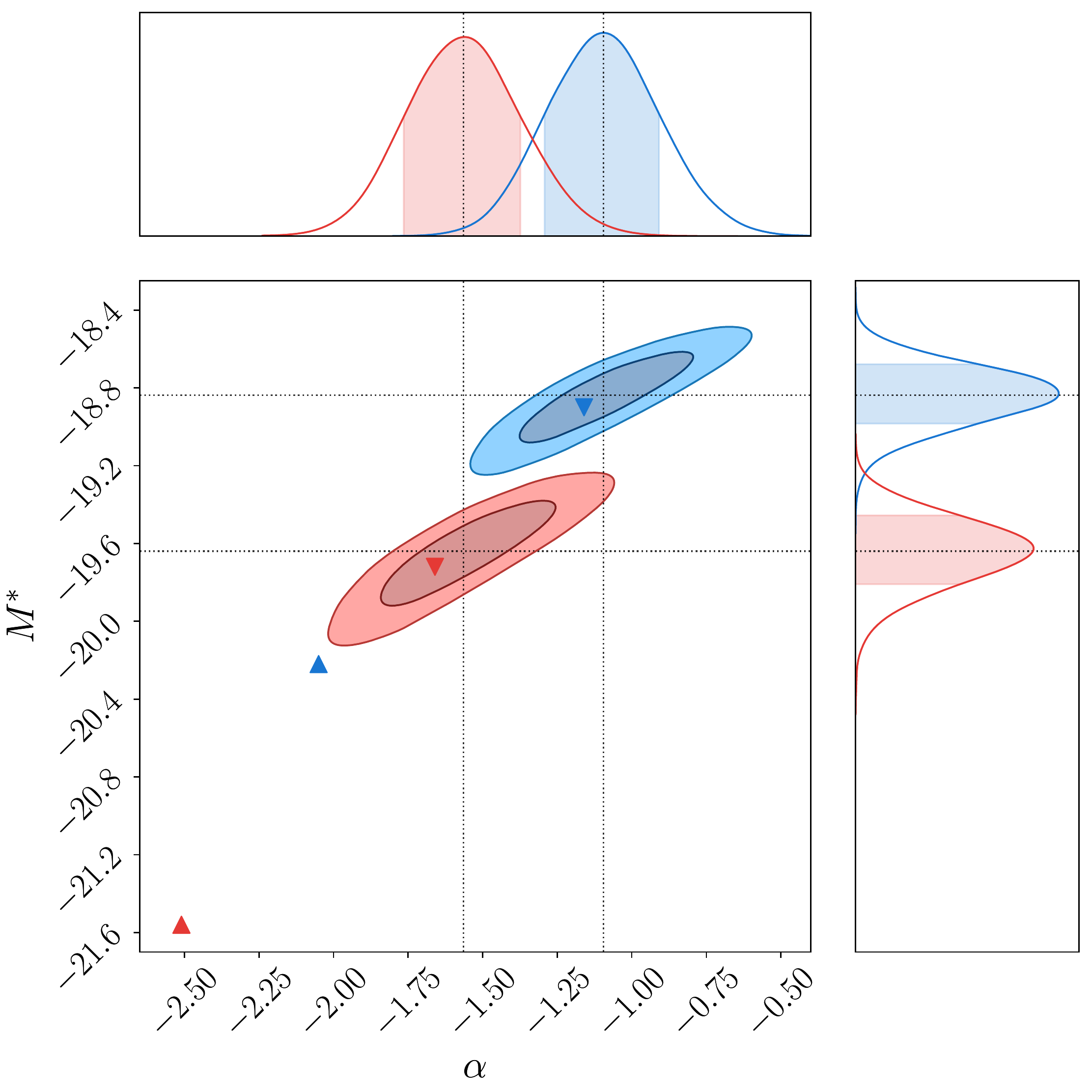}
    \caption{This figure is the same as Fig. \ref{fig:ci}, except that we also plot
    the values tabulated in Table \ref{tab:agn}. The coloured contours have the
    same
    meaning as in Fig. \ref{fig:ci}, with the dotted lines representing values calculated in this work. The upward triangles represent values
    obtained
    by including all the AGN in the sample and the downward triangles show the
    parameters obtained with a sample including no AGNs at all. The red triangles correspond to the red contour representing the redshift bin $0.6 \leq z < 0.8$ and the ones in blue color representing the bin $0.8 \leq z < 1.2$.}
    \label{fig:ci_agn}
\end{figure}

Here we look at the effects on the results if :
1) We remove all the identified X-ray sources, which would mean 
removing more AGN at the expense of excluding some star-forming galaxies without an AGN as well. This case in our survey effectively brings the X-ray luminosity cut down to $\sim 10^{41}$ ergs s$^{-1}$.
2) We do not remove any of the identified X-ray source, 
and compare these two cases with our actual results which have been 
obtained 
by removing only the bright X-ray sources using a luminosity cut.
In table \ref{tab:agn}, we have the values of faint-end slope and characteristic 
magnitude for both these cases, and in Fig. \ref{fig:ci_agn} we plot these parameter
values to compare with our estimates based on a sample with only 
bright X-ray sources removed. 

It is apparent from Fig. \ref{fig:ci_agn} that the small amount of
perturbation in the final results are within the $1 \sigma$ error region if all the sources which are potential AGN, are removed from the sample. 
However, if all these sources are included, the values of both the parameters namely faint-end
slope and characteristic magnitude change by a large margin. 
The $M^{*}$
value brightens by more than a magnitude for $z =$ 0.7, and 
by almost 2 magnitudes for average redshift $z=$ 1, as compared to our
final results for $M^{*}$ tabulated in table \ref{tab:par_f}. 
This is expected,
as there are sources with UV absolute magnitudes reaching $-24$ mag as mentioned in the first paragraph (Fig. \ref{fig:agn}). These very
bright sources affect the bright end of the LF where there are very few or
no galaxies at all.
The brighter values of characteristic 
magnitude drag the faint end slope to extremely steep numbers because
of a strong correlation between these two parameters. 

So we conclude that the bright AGN, if not treated properly can impact the
final UV LF parameters. Their absence from the sample still does not change
the parameters significantly with respect to those obtained if the X-ray
luminosity cut is applied to the source-list removing only the 
bright X-ray sources. 
But, their presence, even in small numbers can 
significantly change the estimated parameters and distort the shape of the UV LF.

\begin{table}
  \setlength{\tabcolsep}{15pt}
  \centering
  \caption{The best-fit values of the fitted Schechter function 
  parameters for both redshift bins in our study, for the two 
  cases discussed in the text.}
  \label{tab:agn}
  \begin{tabular}{lcr}
    \hline\hline
    \noalign{\vskip 0.5mm}
    $\langle z\rangle$ &
    \multicolumn{1}{c}{$M^*$} &
    \multicolumn{1}{c}{$\alpha$} \\
    \hline
    \noalign{\vskip 0.5mm}
    \multicolumn{1}{@{}l}{All AGN included}\\
    0.7$^a$ & $-20.22$ & $-2.05$ \\
    \noalign{\vskip 0.5mm}
    1.0$^a$ & $-21.56$ & $-2.51$ \\
    \hline
    \noalign{\vskip 0.5mm}
    \multicolumn{1}{@{}l}{All AGN removed}\\
    0.7$^a$ & $-18.90$ & $-1.16$ \\
    \noalign{\vskip 0.5mm}
    1.0$^a$ & $-19.72$ & $-1.65$ \\
    \noalign{\vskip 0.5mm}
    \hline
  \end{tabular}\\
  \begin{minipage}{6.5cm}
      \raggedright
      \textsuperscript{$a$}{Average redshift for the bin.}
  \end{minipage}
\end{table}

\bsp
\label{lastpage}
\end{document}